\documentclass{aa_1}  

\usepackage{graphicx}

\usepackage{txfonts}
\usepackage{color}
\usepackage{lscape}
\usepackage{pdflscape}
\usepackage{longtable}

\usepackage{geometry}
\usepackage{booktabs}
\usepackage{multirow}

\usepackage[colorlinks = true,
            linkcolor = blue,
            urlcolor  = blue,
            citecolor = blue,
            anchorcolor = blue]{hyperref}
\usepackage{soul}
\usepackage{comment}

\newcommand{\mstar}{\mbox{$M_{*}$}}
\newcommand{\msun}{\mbox{$M_{\odot}$}}

\newcommand{\zspec}{$z_{\rm spec}$}

\newcommand{\sigmasfr}{$\Sigma_{\rm SFR}$}
\newcommand{\gala}{\texttt{GALAPAGOS-2}}

\newcommand{\idgala}{id$_{\rm GALA}$}
\newcommand{\reff}{$R_{\rm eff}$}

\begin{document} 

   \title{Mass--size evolution and the emerging passive--density relation revealed by JWST/NIRCam in the Spiderweb protocluster}
   \titlerunning{MSR in the Spiderweb protocluster}
   
\author{J. Nadolny\inst{1,2,3}\corrauth{jakub.nadolny@amu.edu.pl} \and J. M. P\'erez-Mart\'inez\inst{2,3} \and H. Dannerbauer \inst{2,3} \and B. Haussler \inst{4} \and M. Huertas-Company \inst{2,3}
\and K. Daikuhara \inst{5} 
\and T. Kodama \inst{6} 
\and Y. Koyama \inst{7,8}  
\and A. Naufal \inst{9} 
\and  P. G. P\'erez-Gonz\'alez \inst{10}
\and Y. H. Zhang \inst{11,12}
}

   \institute{Astronomical Observatory Institute, Faculty of Physics and Astronomy, Adam Mickiewicz University, ul.~S{\l}oneczna 36, 60-286 Pozna{\'n}, Poland 
    \and Instituto de Astrof\'isica de Canarias, E-38205 La Laguna, Tenerife, Spain 
    \and Universidad de La Laguna, Dept. Astrof\'isica, E-38206 La Laguna, Tenerife, Spain 
    \and European Southern Observatory, Alonso de Cordova 3107, Casilla 19001, Santiago, Chile  
    \and Institute of Space and Astronautical Science, Japan Aerospace Exploration Agency, 3-1-1, Yoshinodai, Chuou-ku, Sagamihara, Kanagawa 252-5210, Japan 
    \and Astronomical Institute, Tohoku University, 6-3, Aramaki, Aoba, Sendai, Miyagi 980-8578, Japan 
    \and National Astronomical Observatory of Japan, 2-21-1 Osawa, Mitaka, Tokyo 181-8588, Japan 
    \and  Department of Astronomical Science, The Graduate University for Advanced Studies, 2-21-1 Osawa, Mitaka, Tokyo 181-8588, Japan 
    \and Academia Sinica Institute of Astronomy and Astrophysics (ASIAA) 11F of Astronomy-Mathematics Building, AS/NTU, No. 1, Section 4, 12 Roosevelt Road, Taipei 106319, Taiwan
    \and  Centro de Astrobiolog\'ia (CAB), CSIC-INTA, Ctra. de Ajalvir km 4, Torrej\'on de Ardoz, E-28850, Madrid, Spain 
    \and School of Astronomy and Space Science, Nanjing University, Nanjing, Jiangsu 210093, China 
    \and Key Laboratory of Modern Astronomy and Astrophysics (Nanjing University), Ministry of Education, Nanjing 210093, China 
             }
             
   \date{Received ---; accepted ---}

  \abstract%
  {We investigate how the environment affects galaxy structure in the Spiderweb protocluster at $z=2.16$ using JWST/NIRCam F115W, F182M, and F410M imaging (rest-frame $\sim 3500$\,\AA\ to $1.4\,\mu$m). We perform homogeneous multi-wavelength parametric modelling of the single-S\'ersic and bulge--disc decomposition for the Spiderweb member sample of 103 galaxies within the JWST field of view (up to $\sim 2\times R_{200}$). Star-forming galaxies follow a mass--size relation broadly consistent with the field, with a mildly steeper trend at the high-mass end within uncertainties. Passive galaxies exhibit a flatter mass--size relation, and their typical size (intercept at fixed stellar mass) lies between that of the coeval field and cluster passive populations, indicating an intermediate evolutionary stage. In both star-forming and passive systems, bulges are systematically more compact than discs. Galaxy sizes decrease slightly with increasing wavelength, whereas ALMA-detected dusty star-forming galaxies exhibit a much steeper wavelength dependence, consistent with centrally concentrated, obscured star formation. {  The passive fraction depends primarily on local density: it resembles the field at $\Sigma \lesssim 100$--$200\,{\rm gal\,Mpc^{-2}}$ and rises to $\sim60\%$ at $\Sigma \gtrsim 1000\,{\rm gal\,Mpc^{-2}}$, although with relatively large uncertainties due to the small number of galaxies in the highest-density bins, with no significant additional dependence on clustercentric distance.} We also find a weak but significant correlation between local density and S\'ersic index (strongest in F410M), but no clear correlation with effective radius or the star-formation rate surface density $\Sigma_{\rm SFR}$. These results support a picture in which quenching and structural transformation are already advanced, while the size growth is still ongoing.}


   \maketitle
\nolinenumbers

\section{Introduction}
\label{sec:intro}

The advent of the \textit{James Webb Space Telescope} (JWST) has advanced our understanding of galaxy structure and evolution at redshift $z>1$ \citep{Kartaltepe2023ApJ...946L..15K,HC_CEERS_2024A&A...685A..48H,Shuntov2025arXiv250603243S} up to $z\sim16$ \citep{Ono2025arXiv250208885O}. Before JWST, the high-resolution {\it Hubble Space Telescope} (HST) could observe only the rest-frame UV at $z>3$, missing the optical rest-frame wavelengths \citep[e.g.][]{Oesch2018ApJ,Bouwens2022ApJ}. The ability of JWST to probe these UV--optical rest-frame light with unprecedented depth and resolution has provided insight into some long-standing paradigms. 

In particular, population studies of the stellar mass--size relation (MSR) show that its overall slope and intrinsic scatter remain broadly consistent with pre-JWST studies \citep{vanderwel2014,Mowla2019ApJ...880...57M,Nadolny2021A&A...647A..89N,Nedkova2021MNRAS.506..928N,Nedkova2024MNRAS.532.3747N}: star-forming (SFGs) and passive galaxies follow distinct relations, and average sizes decrease with redshift at fixed stellar mass at different rates for SFGs and QGs \citep{Martorano2024ApJ...972..134M, Ormerod2024,vdW2024ApJ...960...53V,Ward2024ApJ...962..176W,McGrath2026ApJ...999L...6M}, as also shown in simulations \citep[e.g.,][]{Costantin2023ApJ...946...71C}. Furthermore, galaxy sizes vary significantly with rest-frame wavelength \citep{Vulcani2014MNRAS.441.1340V}. For star-forming galaxies, this variation strongly depends on stellar mass \citep{Cheng2024ApJ...977..165J}, whereas sizes of passive galaxies also depend on stellar age, particularly for the most massive objects \citep{Ji2024arXiv,Kawinwanichakij2025}. 

While the MSR reflects the internal processes that determine a galaxy’s size at fixed stellar mass, the morphology--density relation (MDR) captures the influence of the external environment on galaxy morphology, as reflected in the passive--density relation. The MDR has been observed at low \citep{Oemler1974ApJ...194....1O,Dressler1980ApJ...236..351D,Zeleke2019MNRAS.485.1528A,Vulcani2023ApJ...949...73V}, intermediate \citep{Postman2005ApJ,Mei2023A&A...670A..58M} and high redshift \citep{Fudamoto2025arXiv}. The MDR indicates that the relative fraction of different morphological classes varies with the environment. Furthermore, recent studies of low-$z$ clusters show that the fraction of early-type galaxies depends more strongly on the local environment (i.e., distance to the $N^{\rm th}$ neighbour), while late-type (including S0) morphologies depend on the global environment within the cluster (i.e., the clustercentric distance $R/R_{200}$; \citealt{Vulcani2023ApJ...949...73V}). This has also been suggested in clusters and protocusters at higher-$z$ \citep{Mei2023A&A...670A..58M,Pérez-Martínez2023MNRAS.518.1707P,DD_Shi2024ApJ...963...21S}.

With JWST, it is now possible to connect two complementary views of galaxy evolution---structural scaling relations and environmental transformation---at the epoch when clusters are still assembling \citep{Kartaltepe2023ApJ...946L..15K,HC_CEERS_2024A&A...685A..48H,Shuntov2025arXiv250603243S,McGrath2026ApJ...999L...6M}. In protoclusters at $z\gtrsim2$, galaxies span a broad range of conditions, from field-like outskirts to group-like substructures and dense forming cores, making these systems ideal for testing when and where departures from field scaling relations emerge \citep{Muldrew2015MNRAS.452.2528M,Overzier2016A&ARv,Alberts2022Univ....8..554A}. This framework enables direct tests of whether environmental quenching and morphological transformation are already accompanied by measurable structural signatures, such as offsets in the passive MSR and wavelength-dependent size trends linked to dust-obscured growth and bulge build-up \citep{Mei2023A&A...670A..58M,DD_Shi2024ApJ...963...21S,Vulcani2023ApJ...949...73V,Zhang2025A&A}. JWST/NIRCam provides the required rest-frame optical resolution and depth for homogeneous multi-band structural modelling across these environments \citep{Kartaltepe2023ApJ...946L..15K,HC_CEERS_2024A&A...685A..48H}.

Protoclusters are large-scale, massive overdensities---typically spanning tens to hundreds of cubic megaparsecs---that trace the nodes of the cosmic web \citep{Bond1996Natur.380..603B} and are found at and before the peak of the cosmic star formation rate (SFR) density at $z>2$ \citep{Muldrew2015MNRAS.452.2528M,Overzier2016A&ARv,Alberts2022Univ....8..554A} and show evidence for an emerging intra-cluster medium (\citealt{DiMascolo2023Natur.615..809D,Lepore2024A&A...682A.186L} in Spiderweb).
Observationally, protoclusters are often identified around massive signposts such as quasars, dusty star-forming galaxies, and radio-loud galaxies at $z>2$ \citep[e.g.,][]{Carilli1997,Wylezalek2013ApJ...769...79W,HD2014A&A,Calvi2023A&A}, with recent discoveries extending to $z>5$ \citep{Morishita2023ApJ...947L..24M,Wang2025NatAs,Galbiati2025A&A...696A..95G,sun2025bigfoot,akins2025jwstalma,Fudamoto2025arXiv,LiQiong_2025,Morishita2025ApJ...983..152M}, dominated by measurments using JWST. For example, \citet{sun2025bigfoot} identified the Bigfoot structure ($z\sim4$) comprising 11 subgroups, where the most massive subgroups exhibit an enhanced fraction of massive galaxies relative to lower-mass haloes and the field; at $z\sim7.5$ \citet{Fudamoto2025arXiv} found that a protocluster core environment can drive rapid and efficient formation and evolution of massive galaxies at very young cosmic times. Similarly, \citet{LiQiong_2025} reported 26 overdensity candidates at $5<z<7$ in JWST wide-field surveys, finding no significant environmental dependence of the MSR, but slightly elevated star formation rates for galaxies in overdense regions compared to the field.

In this work, we study the MSR and passive--density relation of the Spiderweb protocluster. The radio galaxy PKS1138-262 \citep[][hereafter the Spiderweb galaxy]{Carilli1997} at $z=2.16$ is the central galaxy of the Spiderweb protocluster \citep{Pentericci1997A&A...326..580P,Kurk2000A&A...358L...1K,Pentericci2000}. Since its discovery, many authors have studied this overdensity using data spanning the entire electromagnetic spectrum. Using a selection of samples, such as H$\alpha$ emitters (HAE), submillimeter galaxies (SMGs), or X-ray emitters \citep{Kodama2007MNRAS.377.1717K,HD2014A&A,Pérez-Martínez2023MNRAS.518.1707P,Shimakawa2024,Tozzi2022A&A...662A..54T,Perez-Martinez_2024}, key findings have been reported. These are the following: the emergence of the red sequence \citep{Kurk2004A&A...428..817K,Kodama2007MNRAS.377.1717K,Zirm2008ApJ...680..224Z}, the dust and gas content of protocluster members \citep{Emonts2018MNRAS.477L..60E,Tadaki2019PASJ...71...40T,Jin2021A&A...652A..11J,Perez-Martinez-COALASIII2025A&A}, the signs of environmental effects on SFRs, gas properties, MSR \citep{Pérez-Martínez2023MNRAS.518.1707P,Kazuki2024MNRAS.531.2335D}, the X-ray influence on the SFRs \citep{Shimakawa2024,Shimakawa2025_Xray}, or the inside-out quenching mechanisms that correlates with morphology \citep{Laishram2026ApJ...998..158L}. Several authors also studied structural parameters of the SW protocluster member galaxies using ground-based observations with moderate resolution \citep{Zirm2012ApJ...744..181Z,Pérez-Martínez2023MNRAS.518.1707P} or space-based HST data \citep{Koyama2013MNRAS.428.1551K,Naufal2023ApJ...958..170N} limited to the UV rest-frame wavelengths. \citet{Pérez-Martínez2023MNRAS.518.1707P} have shown the MSR of the selected HAEs in Spiderweb, finding that this population lies near the MSR for the field LTGs at a similar redshift to that given by \citet{vanderwel2014}, with a few galaxies falling on the relation for the ETGs. The size estimates are based on the VLT/HAWKI instrument in K-s band with a limiting \reff\ of 1.7 kpc. 
Recently, \citet{Zhang2024A&A...692A..22Z} showed that ALMA-detected galaxies in Spiderweb trace a DSFG overdensity about twice that in the field, with strong clumping in the eastern outskirts, where local excesses reach $\gtrsim10$ times the field value and are offset from the HAE peak. Using JWST/NIRCam, \citet{Zhang2025A&A} analysed ten of these ALMA-detected DSFGs and found unusually large stellar discs, consistent with accelerated structural growth in dense, infalling substructures, in line with the filamentary-accretion scenario of \citet{Shimakawa2018MNRAS.481.5630S}. In this work, we complement ALMA-focused studies by performing a homogeneous structural analysis of all accessible Spiderweb member galaxies within the JWST FoV, thereby placing dusty systems in the context of the full protocluster population.

This paper is organised as follows. In Section~\ref{sec:data} we describe the JWST/NIRCam data and the construction of the Spiderweb protocluster morphological sample within the JWST field of view (up to $\sim2\times R_{200}$). In Section~\ref{sec:methods} we present our multiwavelength parametric modelling of galaxy light profiles, including single-component S\'ersic fits and bulge--disc decompositions, together with the derived structural parameters. In Section~\ref{sec:results} we present the resulting mass--size and passive--density relations, and in Section~\ref{sec:discussion} we discuss their implications in the broader context of field and (proto)cluster galaxy evolution.
Throughout this paper, we assume \citet{Chabrier2003PASP..115..763C} initial mass function and a cosmological model with
$h$ = 0.676, $\Omega_\Lambda = 0.685$, $\Omega_m = 0.315$ \citep{Planck2018}.


\section{Data}
\label{sec:data}
\subsection{JWST NIRCam imaging}
\label{sec:data_jwst}
This work uses JWST NIRCam imaging in the F115W, F182M, and F410M filters from Cycle 1 GO 1572 (PI: Dannerbauer and Koyama). The corresponding rest-frame wavelength are $\sim3600$, $5800$, and $12900\,\mathrm{\AA}$.  Details of the observations, reduction, and mosaic assembly are provided in \citet{Shimakawa2024ApJ...977...73S} and \citet{Perez-Martinez_2024}. The F182M PSF full-width half-maximum is 0.06 arcsec \footnote{\url{https://jwst-docs.stsci.edu}}, corresponding to $\sim0.25$ kpc at $z=2.16$. The field of view covers the central part of the Spiderweb protocluster to an extent twice that of $R_{200}$ (see Fig. \ref{fig:ra_dec}). 

\begin{center}

\begin{table*}
\centering
\caption{Galaxy properties}
\label{tab:properties}
\begin{tabular}{lllllllll}
\toprule
 ID &  GALA ID &  MEMBER & $z_{\rm spec}$ &                                $z_{\rm spec}$ ref. & M$_* \times 10^{10}$ & M$_*$ ref. &               SFR & SFR ref. \\
\midrule
 16 &       39 &       1 &          2.144 & [4] &      $4.11 \pm 0.91$ &        [4] &  $38.44 \pm 6.27$ &      [4] \\
 17 &      119 &       1 &          2.163 & [4] &      $1.01 \pm 0.17$ &        [4] &  $44.74 \pm 2.24$ &      [4] \\
 21 &     1352 &       1 &          2.161 & [4] &      $0.72 \pm 0.23$ &        [4] &  $22.04 \pm 5.21$ &      [4] \\
 22 &     1388 &       1 &          2.163 & [4] &      $1.42 \pm 0.32$ &        [4] & $62.33 \pm 11.38$ &      [4] \\
 23 &     1381 &       1 &          2.162 & [4] &      $0.83 \pm 0.13$ &        [4] &  $34.94 \pm 1.93$ &      [4] \\
 24 &      272 &       1 &          2.163 & [4] &      $0.49 \pm 0.07$ &        [4] &  $21.07 \pm 2.71$ &      [4] \\
 26 &      273 &       1 &          2.163 & [4] &      $0.48 \pm 0.07$ &        [4] &  $21.34 \pm 2.43$ &      [4] \\
 27 &      308 &       1 &          2.158 & [4] &     $23.23 \pm 2.92$ &        [4] &   $2.54 \pm 1.23$ &      [4] \\
 28 &      337 &       1 &          2.162 & [4] &      $9.19 \pm 2.00$ &        [4] &   $5.33 \pm 3.43$ &      [4] \\
 37 &     2007 &       1 &          2.164 & [4] &      $0.27 \pm 0.07$ &        [4] &  $12.43 \pm 3.29$ &      [4] \\
\bottomrule
\end{tabular}

\vspace{1ex}
\parbox{\linewidth}{\footnotesize\textbf{Note.} First 10 rows of the catalogue of objects studied in this work. 
ID: unique ID for the overall protocluster members. GALA ID is the unique ID from \gala. 
MEMBER is the protocluster membership flag: 1 to 5, as described in Section \ref{sec:Sample}. 
Note that for MEMBER > 2, $z_{spec}$ is assumed to be $2.16$. 
The spectroscopic redshift references column: 1: \citet{Koyama2013MNRAS.428.1551K}; 2: \citet{Kurk2004A&A...428..817K}; 
3: \citet{Perez-Martinez_2024}; 4: \citet{Shimakawa2024}; 5: \citet{Shimakawa2024ApJ...977...73S}; 
6: \citet{Pentericci2000}; 7: \citet{Kuiper2011MNRAS.415.2245K}; 8: \citet{Croft2005AJ....130..867C}; 9: \citet{Doherty2010A&A...509A..83D}; 
10: \citet{Tanaka2013ApJ...772..113T}; 11: \citet{Naufal2024ApJ...977...58N}; 12: \citet{HD2014A&A}; 13: \citet{Emonts2018MNRAS.477L..60E}; 
14: \citet{Tadaki2019PASJ...71...40T}; 15: \citet{Jin2021A&A...652A..11J}; 16: \citet{Zhang2024A&A...692A..22Z}; 17: \citet{Tozzi2022A&A...662A..54T}. 
Multiple references are given in an electronic version of this table.
}
\end{table*}
\end{center}

\subsection{Sample selection and complementary data}
\label{sec:Sample}

\begin{figure*}[t!]
\includegraphics[width=0.5\textwidth,clip]{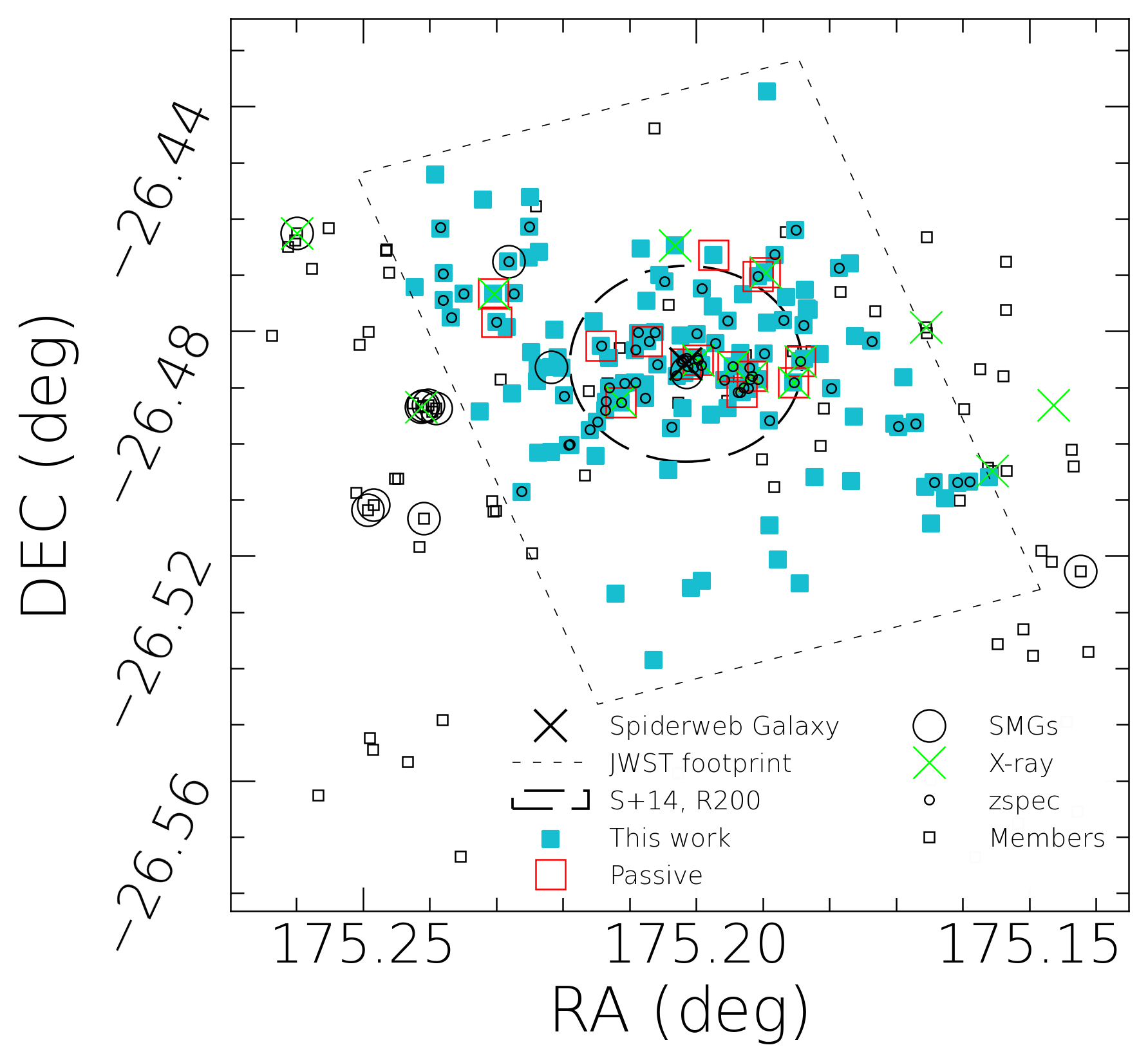}
\includegraphics[width=0.5\textwidth,clip]{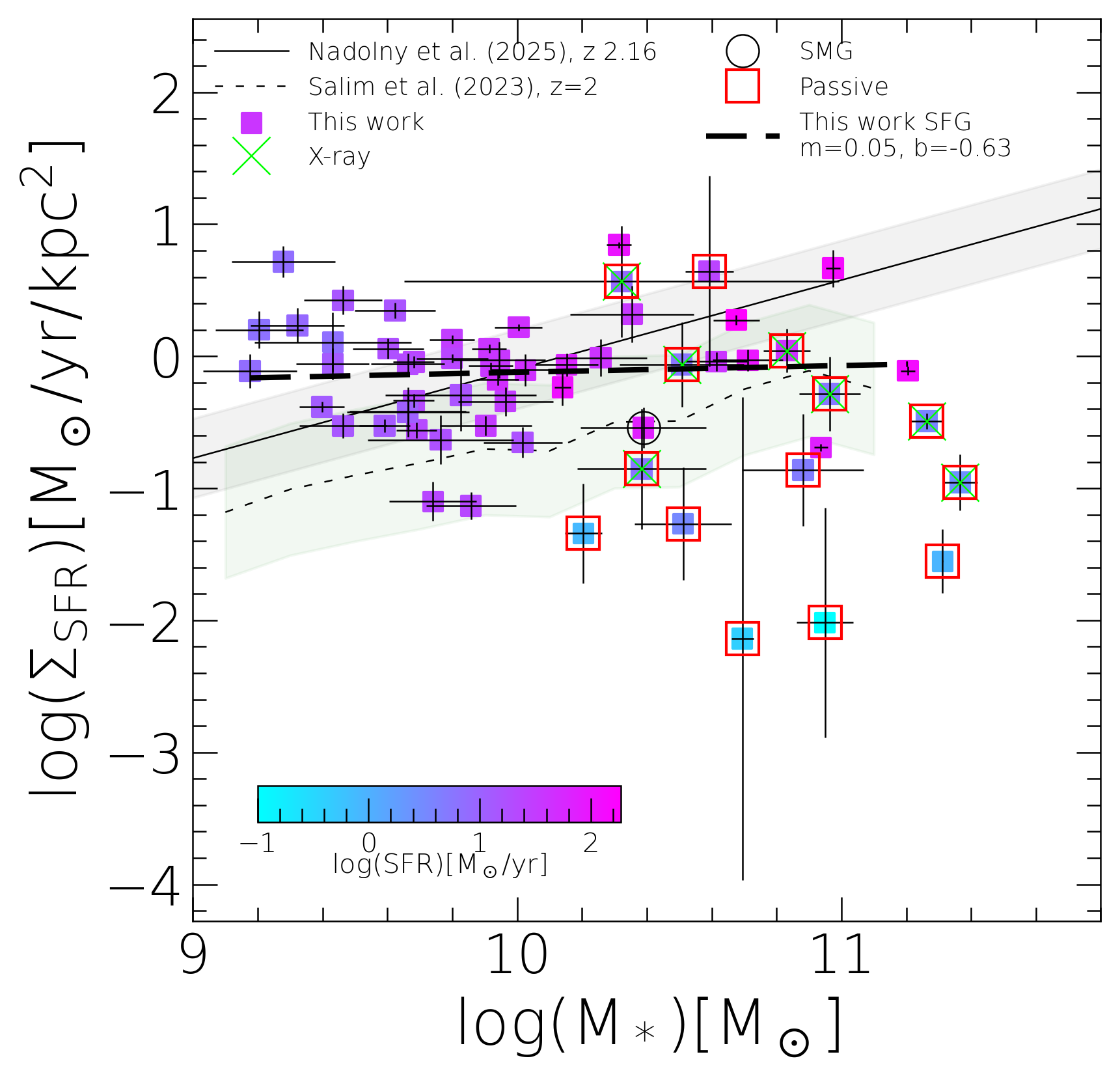}
    \caption{(Left) Spatial distribution of Spiderweb galaxies. {  Cyan squares show our parent sample, while smaller black empty squares show the rest of the galaxies in the Spiderweb protocluster, and empty small circles show member galaxies with spectroscopic confirmation.} Black empty circles and squares denote SMGs from \citet{HD2014A&A}, and passive galaxies selected using the SFMS, while the green X marker denotes X-ray-detected galaxies from \citet{Tozzi2022A&A...662A..54T}. Black circle denotes $R_{200}$ from \citet{Shimakawa2014MNRAS.441L...1S}. (Right) Star formation rate surface density \sigmasfr\ as a function of stellar mass, colour-coded by SFR. The \sigmasfr\ main sequences are from:  \citet[][simulations at z=2.16]{Nadolny2025A&A}, \citet[][observations, at $z\sim2$]{Salim2023ApJ...958..183S}, and from this work for SFGs. 
    \label{fig:ra_dec}}
\end{figure*}

The primary goal of this work is to provide a parametric morphological analysis of all the Spiderweb protocluster galaxies within the JWST footprint shown in Figure \ref{fig:ra_dec}, regardless of their nature. In what follows, we provide a list of objects sorted by their membership status. (MEMBER 1) The innermost member galaxies, defined based on \zspec\ as galaxies with $2.14 < $\zspec $< 2.18$, comprise 107 sources. (2) Infalling member, defined as galaxies with  $2.09 < $\zspec $< 2.23$ with a total of 26 sources, that are not in (1). (3) Simultaneous HAEs and PBEs without \zspec\ confirmation, with a total of 5 sources. (4) Only HAEs without \zspec\ (total of 40 sources), and (5) only PBEs without \zspec\ (total of 22 sources). Whenever the galaxy has \zspec, we use it; for the rest (confirmed photometrically as HAEs or PBEs, $\sim30\%$ of the final sample), we assume a median redshift of the Spiderweb protocluster ($z=2.16$). These objects were recovered from the following works. Objects detected as HAEs: \citet{Kurk2004A&A...428..817K,Koyama2013MNRAS.428.1551K,Shimakawa2024,Pérez-Martínez2023MNRAS.518.1707P}; PBEs \citet{Shimakawa2025_Xray}; optical emission line galaxies \citet{Kuiper2011MNRAS.415.2245K}; extremely red objects \citet{Croft2005AJ....130..867C,Doherty2010A&A...509A..83D,Tanaka2013ApJ...772..113T,Naufal2024ApJ...977...58N}; sub-millimeter galaxies (SMGs): \citet{HD2014A&A,Emonts2018MNRAS.477L..60E,Tadaki2019PASJ...71...40T,Jin2021A&A...652A..11J,Zhang2024A&A...692A..22Z}; X-ray detected sources from \citet{Tozzi2022A&A...662A..54T}. Finally, we also study three objects from \citet{Zhang2025A&A} not included in \citet{Zhang2024A&A...692A..22Z}. A total of 263 unique sources associated with the Spiderweb protocluster were retrieved from the literature (136 with \zspec). Of these, 140 objects fall within the JWST footprint (80 with \zspec), and these constitute our parent sample. Assuming a $\sim10\%$ interloper rate for the narrow-band selection \citep[][i.e. out of 60 objects without \zspec]{Sobral2013MNRAS.428.1128S,Pérez-Martínez2023MNRAS.518.1707P}, we expect $\sim6$ interlopers in this subsample, corresponding to an overall contamination of $\sim4\%$ (6/140) in the final morphological sample. For 47\% (67/140) of the sample with measured SFRs, using the star formation main sequence (SFMS) from \citet{Speagle2014ApJS..214...15S} we divided the sample into galaxies on the SFMS (52/67, 78\%) and galaxies below SFMS (15/67, 22\%), the latter lying at least $2$ times below the SFMS at the given redshift {  (or with log(sSFR)$<-9.2$yr)}. While this is not a strict definition of passive and active galaxies, for convenience, we will refer to these as star-forming (SFGs) and passive galaxies. Our parent sample is shown in Figure \ref{fig:ra_dec}, together with galaxies that were not detected or are outside of the JWST FoV (black empty squares). We also indicate passive, SMGs, or X-ray-detected galaxies by different markers, as explained in the legend. 

Stellar masses and SFRs reported in this work were retrieved from \citet{Shimakawa2024}, \citet{Pérez-Martínez2023MNRAS.518.1707P}, \citet{Tanaka2013ApJ...772..113T}, \citet{Naufal2024ApJ...977...58N}, and \citet{Zhang2024A&A...692A..22Z}, with priority given to \citet{Shimakawa2024}. Whenever necessary, the masses and SFR were recalibrated assuming a \citet{Chabrier2003PASP..115..763C} initial mass function, and \citet{BruzualCharlot2003} stellar population synthesis models. Further sample refinements are given in Section \ref{sec:methods_parametric}.


\subsection{Comparison data}
\label{sec:comparision_data}
We use several works based on HST and/or JWST to place our measurements in the broader context of the mass--size and passive--density relations. 
Regarding the MSR, we use the COSMOS2025 catalogue recently presented in \citet{Shuntov2025arXiv250603243S} based on the COMOS-Web JWST imaging data. Using this catalogue, we obtain the MSR for the field galaxies with redshift between 2 and 2.4, and with SFG and passive selection based on the rest-frame NUV--r--J colour distribution \citep{Ilbert2013A&A...556A..55I}. See \citet{Shuntov2025arXiv250603243S} for details about their methods. Similarly, we use JWST data presented in \citet{Martorano2024ApJ...972..134M} within the same redshift range, using U--V--J colour to select star-forming and passive galaxies. We also show \citet{vanderwel2014} HST-based MSR for comparison with pre-JWST works. In our analysis, we will use the typical size (or intercept) from \citet{vanderwel2014,Mowla2019ApJ...880...57M,Afanasiev2023A&A...670A..95A,Ward2024ApJ...962..176W}. The common way to measure the typical size is through the mass--size relation intercept evaluated at the pivotal stellar mass of $5\times10^{10}$\msun. For the passive--density relation, we use clusters from the CARLA survey ($z > 1.9$) from \citet[][measurements from their table 5]{Mei2023A&A...670A..58M}, based on HST imaging, where passive galaxies are selected using the UVJ diagram. Our passive sample is defined as galaxies below the SFMS; although the two criteria are not identical, both are designed to isolate low-sSFR systems, so they provide a suitable first-order comparison of passive-fraction trends with environment. However, as demonstrated by \citet{Nedkova2021MNRAS.506..928N}, the resulting MSR for passive galaxies does not depend on how quiescence is defined, whether through star-formation activity, rest-frame colours, or structural properties.

\section{Methods}
\label{sec:methods}
\subsection{Parametric modelling}
\label{sec:methods_parametric}

In this work we use \gala\ \citep{Barden2012MNRAS.422..449B,Haussler2013MNRAS.430..330H,Haussler2022A&A...664A..92H}, a wrapper software for SExtractor \citep{Bertin1996A&AS..117..393B} and GLAFIT-M to perform single S\'ersic (SS) modelling and bulge--disc (BD) decomposition. Details of how \gala\ works can be found in \citet{Haussler2013MNRAS.430..330H} and references therein. A short and essential description follows.

\gala\ use SExtractor to build the source catalogue on the selected detection image in two separate runs, so-called high dynamical range (HDR) mode. First, the "cold" run is designed to detect bright (and usually large) sources without splitting them into several sources, while the second, the "hot" run, is optimised to detect faint sources.  The two catalogues (cold and hot) are combined internally to provide one single detection catalogue and a combined segmentation map. As a detection image, we used the NIRCam F182M filter corresponding to $\sim 5800$\AA\ rest frame, the wavelength that has been traditionally used for MSR measurements \citep[][to name a few]{vanderwel2014,Mowla2019ApJ...880...57M,Nadolny2021A&A...647A..89N,Nedkova2021MNRAS.506..928N,Ward2024ApJ...962..176W}

The combined catalogue is used to provide input parameters for GALFIT-M \citep{Haussler2013MNRAS.430..330H,Vika2014MNRAS.444.3603V}, which uses Chebyshev polynomials to fit the selected model to all given bands simultaneously. This is particularly important for low-SN bands because one can obtain meaningful results for these filters \citep{Vika2014MNRAS.444.3603V,Nadolny2021A&A...647A..89N}, which had been proven difficult when using a single-band setup available in GALFIT \citep{Peng2002AJ....124..266P}. Based on the segmentation map and the brightness of the particular galaxy, the decision is made whether the neighbouring source will be fitted as well. In particular, if the neighbouring source is brighter or close enough, this source is fitted together with the target object. Our parametric modelling adopts a \cite{Sersic1968adga.book.....S} surface-brightness profile:
\begin{equation}
\label{eq:sersic}
I (r) = I_{e} \exp \{-b_{n}[(r/r_{e})^{1/n} - 1]\},
\end{equation}
where $r_{e}$ is the effective radius (i.e., the radius containing 50\% of the total flux), $I_{e}$ is the intensity at $r_{e}$, $n$ is the S\'ersic index, and $b_{n}$ is a function of $n$ \citep{Ciotti1991A&A...249...99C}. See Table \ref{tab:sextractor} in Appendix \ref{sec_app:sextractor} for details of relevant parameters used in the SExtractor setup. These particular parameters were found after trial and error to find best results to detect faint and small galaxies while not splitting bigger galaxies into several detections. This strategy has been applied successfully for decades \citep[e.g.][]{Rix_2004,Barden2012_GALAPAGOS,Nadolny2021A&A...647A..89N}. For the fitting purposes we simulated PSFs using the \texttt{STPSF/webbpsf} script \citep{Perrin_WebbPSF} for all the filters.

Additionally to SS, we also perform the BD decomposition. This involves the simultaneous fitting of two components to selected galaxies, allowing for the separation of the central bulge and extended disk. While this modelling is challenging, it has been shown to provide valuable insights on the galaxy evolution \citep{Vika2014MNRAS.444.3603V,Dimauro2019MNRAS.489.4135D}. 

After several tests, we set the same Chebyshev polynomial orders for single S\'ersic (SS) modelling and BD decomposition. We kept constant position, axis ratio, and position angle; S\'ersic index, and effective radius were set to vary linearly with the wavelength; magnitude were set to be free parameter.

Finally, the fitting process is constrained to explore only the physically (or instrumentally) motivated space of parameters. The S\'ersic index is limited between 0.2 and 8, the effective radius is limited between 0.3 and 400 pixels (with pixel scale of 30 mas/px), and magnitudes are not allowed to deviate more than 5 mag from the input magnitude. For details and motivation of these constraints, read Section 3 in \citet{Haussler2013MNRAS.430..330H}.

As a result of the fitting, we obtain the S\'ersic index ($n$), magnitude, effective radius ($R_{\rm eff}$), axis ratio, and position angle for each detected source in each band, as well as for the bulge and disk components. Out of 140 sources in our parent sample, we detected 114 in F182M image.  Finally, we exclude the Spiderweb galaxy, three further objects that are on the border of the FoV, and seven visually selected bad fits, leaving a final 103 galaxies in our morphological sample, out of which 70 have \zspec, and 58 with measured SFR (44 SFGs, and 14 passive galaxies). Furthermore, and only for BD decomposition analysis, we considered only galaxies with measured dik sizes larger than bulge sizes, excluding a total of {  eight} SFGs and one passive galaxy. {  The excluded galaxies do not appear to introduce any obvious systematic effect into the results, as their stellar masses, sizes, S\'ersic indices, and SFRs span the full range of values probed in this study. Therefore, this selection is not expected to bias the BD analysis or affect the SS results.}




\subsection{Star formation rate surface density}
\label{sec:SigmaSFR}

Using single SS effective radii $R_{\rm eff,F182M}$ and available SFRs, we estimate the star formation rate surface density $\Sigma_{\rm SFR}={\rm SFR}/(2\pi R_{\rm eff}^2)$[\msun/yr/kpc$^2$], showed in Figure \ref{fig:ra_dec}, right panel. The \sigmasfr\ has been shown to greatly complement commonly used diagnostic diagrams to differentiate between star-forming and passive galaxies. It has been demonstrated observationally \citep{Salim2023ApJ...958..183S,Calabro2024arXiv240217829C} and in simulations \citep{Nadolny2025A&A} to evolve with redshift. In particular, \citet{Nadolny2025A&A} provide a consistent framework to estimate the relation between \sigmasfr\ and stellar mass at a given redshift (\sigmasfr-MS). In the following, we will use their equation 1, 2 and 3 to estimate the \sigmasfr-MS at the redshift of 2.16.

\subsection{Environment estimation}
\label{sec:methods_environment}
We quantify environment using the projected local galaxy density based on the distance to the third nearest neighbour, $\Sigma_3$, defined as:
\begin{equation}
\Sigma_{N}=\frac{N}{\pi r_{N-1}^{2}},
\end{equation} where $N=3$. 
This method allows for taking into account local density peaks of small groups, which are not necessarily close to the protocluster centre. In this estimation, we include all the confirmed member galaxies of the Spiderweb protocluster, including those outside of the JWST FoV. 

\section{Results}
\label{sec:results}

\subsection{General properties of morphological sample}
\label{sec:results_spatial_distr}


The right panel of Figure \ref{fig:ra_dec} shows \sigmasfr\ as a function of stellar mass, together with the simulated \sigmasfr-MS relation at $z=2.16$ \citep{Nadolny2025A&A} and observed \sigmasfr-MS from \citet{Salim2023ApJ...958..183S} for field galaxies. Most of Spiderweb protocluster galaxies are found around the \sigmasfr-MS from simulations, while slightly above the observed relation from \citet{Salim2023ApJ...958..183S}. The \sigmasfr-MS for Spiderweb protocluster SFGs is flatter, with a slope of 0.05 and an intercept of -0.63. The majority (12/14) of passive galaxies, and all but one X-ray-detected galaxy, are found below the \sigmasfr-MS. As expected, these are the galaxies with the lowest \sigmasfr\ from our sample. 


The percentage of galaxies with S\'ersic index $n<2.5$ is 88, 88,  and 84\%, and the mean values of S\'ersic indices are 1.35, 1.48, and 1.55 for F115W, F182M, and F410M. The general population of Spiderweb galaxies is composed of galaxies described by rather low-$n$ indices, usually found in star-forming late-type galaxies at this redshift with JWST \cite{Ward2024ApJ...962..176W,Martorano2025A&A...694A..76M}. The distribution of S\'ersic indices in all the filters used in this work is shown in the top panel in Figure \ref{fig:ss_index}. 

The bottom panel in Figure \ref{fig:ss_index}, show the S\'ersic index in F182M for SFG and passive galaxies (as described in Sec. \ref{sec:Sample}). Our SFGs subsample gives a mean S\'ersic index of $n_{\rm F182M}=1.1$, while passive galaxies show a higher mean S\'ersic index of $n_{\rm F182M}=2.25$. {  Despite this difference in the mean values, the two distributions overlap substantially, indicating that the passive sample is not composed exclusively of spheroid-dominated systems.} The SFGs S\'ersic index is in good agreement with \citet{Ward2024ApJ...962..176W} and \citet{Martorano2025A&A...694A..76M}. Considering the passive Spiderweb protocluster galaxies, the mean $n$ value is lower as compared to the field passive galaxies \citep[mean $n$ between $2.6$ and $3$, see Fig. 6 in][]{Martorano2025A&A...694A..76M}, or to the passive galaxies found at the centres of the protoclusters at similar redshift \citep[e.g.:][two passive galaxies with $n=4.5$ and $6.3$]{DD_Shi2024ApJ...963...21S}. Objects from \citet{DD_Shi2024ApJ...963...21S} however are among the brightest cluster galaxies, and are on their way to merge at the very centre of their protocluster to became a BCG. In our morphological sample, we have six galaxies with $n>3$, two passive (with \mstar$\sim10^{10.4}$\msun) and one low-mass (\mstar$\sim10^{9.2}$\msun) star-forming, while the rest have no SFR measurements. All these galaxies are found within the $R_{200}$, however, not in the closest vicinity of the Spiderweb galaxy. 


\begin{figure*}[ht!]
\includegraphics[width=0.95\textwidth,clip]{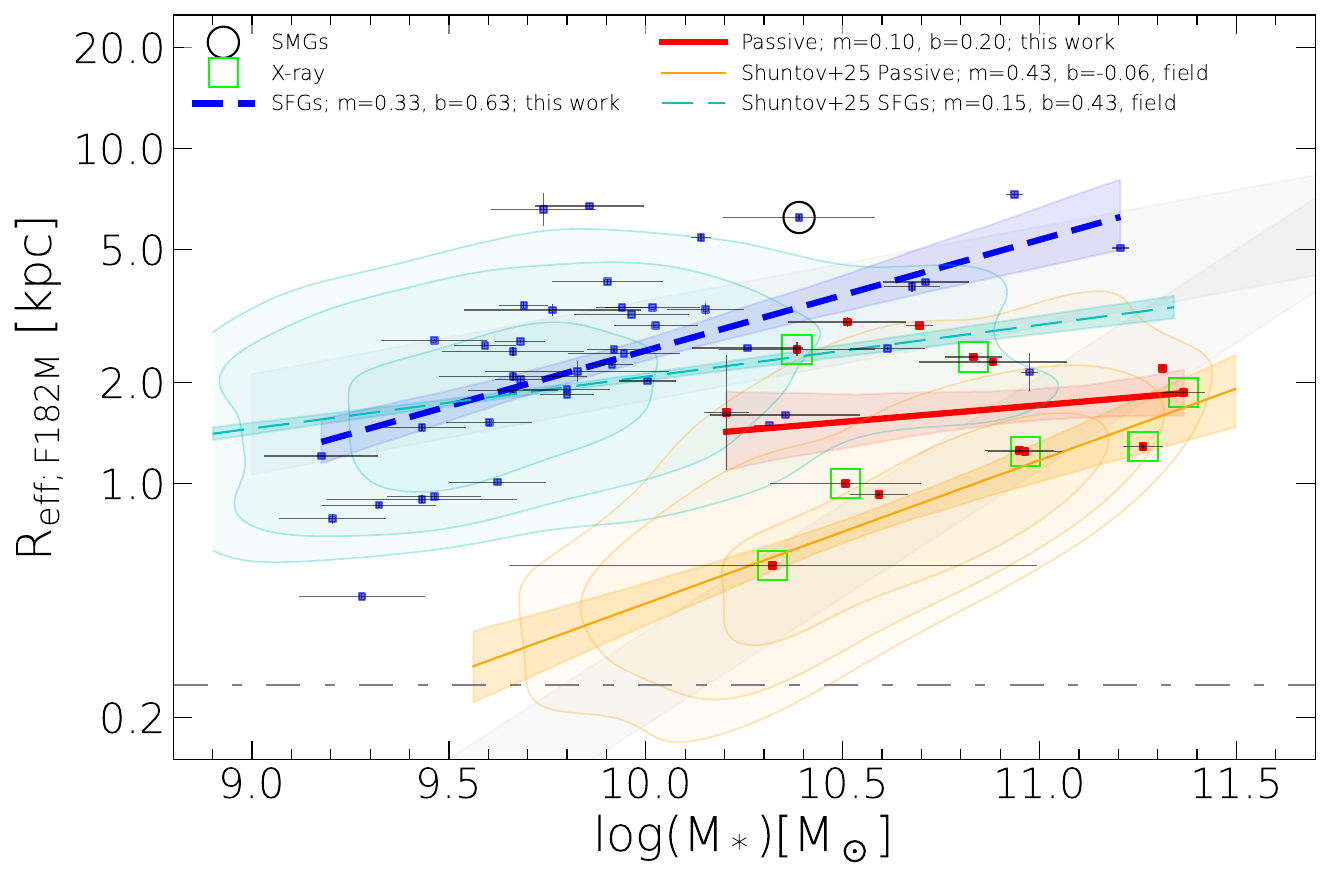}
    \caption{Mass--size relation using sizes based on a single S\'ersic F182M model (rest-frame $\sim5800\,\mathrm{\AA}$). Blue and red squares show SFGs and passive galaxies from the Spiderweb protocluster. Black empty circle and green X markers show SMGs and X-ray galaxies from \citet{HD2014A&A} and \citet{Tozzi2022A&A...662A..54T}. The blue dashed and red solid lines show the relation in the form $log(R_{\rm eff})[{\rm kpc}]= m\times log(M_*) + b$, using SFGs and passive galaxies with the slope and intercept at \mstar=$5\times10^{10}$\msun\ indicated in the legend. The shaded regions represent 68\% confidence intervals estimated after 1\,000 bootstrap resamplings. Grey hatched regions are LTGs and ETGs from \citet{vanderwel2014} at $z=2.25$. Cyan and orange shaded regions are NUV–-r–-J-selected field SFGs and passive galaxies from \citet{Shuntov2025arXiv250603243S} (see Sec. \ref{sec:comparision_data}), with the respective fits in the same colours with slopes and intercepts given in the legend. Passive Spiderweb protocluster MSR is flatter with a larger typical size than the field passive galaxies.
    \label{fig:msr}}
\end{figure*}

\subsection{Mass--size relation}
\label{sec:results_MSR}

In this section, we describe the mass--size relation for the Spiderweb protocluster galaxies in two steps. First, we will show the results of the SS model in F182M (rest-frame $\sim5800\,\mathrm{\AA}$) for subsamples selected based on the relative position to SFMS (SFGs, passive galaxies). This rest-frame wavelength enables an straightforward comparison with the most of the studies regarding MSR. Then, we will consider the results for the remaining filters and for the BD decomposition. 

\subsubsection{Single S\'ersic mass--size relation}
\label{sec:results_SS_MSR} 
Here, we will investigate the effective radius of our SS sample and the corresponding MSRs for the star-forming and passive protocluster populations. We found a mean $R_{\rm eff, F182M}$ of $2.39\pm0.11$ kpc for our morphological sample using the SS model, which includes all the galaxies, while we obtain the mean $R_{\rm eff, F182M}$ of $2.64\pm0.14$ kpc if considering \zspec\ subsample. The mean $R_{\rm eff, F182M}$ for SFGs only is $2.80\pm0.15$ kpc (or $3.18\pm0.10$ kpc for \zspec\ sample). This value is in agreement (within the errors) with the value of $3.14\pm0.21$ kpc from \citet{Pérez-Martínez2023MNRAS.518.1707P}, where a subsample of HAEs was studied. On the other hand, the mean effective radius of $R_{\rm eff, F182M} = 1.79 \pm 0.10$ kpc ($1.70\pm 0.10$ kpc) is estimated for (\zspec\ subsample of) passive galaxies only. 

Using the effective radius of the SS fit to the F182M images, we can derive the mass--size relation parametrised as:
\[\log(R_{\rm eff}/{\rm kpc}) = m \, (\log M_\star - \log(5\times10^{10}M_\odot)) + b,\]
we find the following slopes $m$, intercepts $b$, and dispersion $\sigma$. For SFGs: $m = (0.33\pm0.08)$, $b = (0.62\pm0.07)$, with 0.22 dex dispersion. For passive galaxies:
$m = (0.10\pm0.16)$, $b = (0.20\pm0.06)$, with 0.22 dex dispersion. These results are tabulated in Table \ref{tab:MSR_results}.

As a robustness check against the assumed redshift for sources without \zspec\ (Sec. \ref{sec:Sample}), we repeated the fit using only galaxies with spectroscopic redshifts. The best-fit relations are consistent within the scatter of Fig. \ref{fig:msr}: for SFGs we obtained $m=0.29\pm0.07$, $b=0.66\pm0.06$, and $\sigma=0.19$; for passive galaxies we obtained $m=0.15\pm0.18$, $b=0.18\pm0.08$, and $\sigma=0.21$. Thus, the conclusions regarding the relative location of the Spiderweb MSR with respect to the field are unchanged when restricting the analysis to the \zspec\ subsample. {  We also tested against a more conservative selection of passive galaxies using a limit of 1 dex below SFMS (or with log(sSFR)$<-9.6$ yr). This selection removes four passive galaxies with stellar masses of log(\mstar)$\sim10^{10.6}$\msun (three of them are X-ray detected objects), including them in the SFGs subsample. The overall results of MSR remain unchanged, with only minor changes in the obtained slope and intercept parameters.}

Our MSR for SFGs agrees with the relation for the field SFGs from \citet[][HST]{vanderwel2014} and \citet[][JWST; black and magenta dotted lines in Fig. \ref{fig:msr}]{Ward2024ApJ...962..176W,McGrath2026ApJ...999L...6M}. Using the COSMOS2025 catalogue \citep[][see Sec \ref{sec:comparision_data}]{Shuntov2025arXiv250603243S} we obtained a slightly flatter MSR (with slope $m=0.15$) as compared to Spiderweb SFGs. However, the bulk of the Spiderweb SFGs lie within the COSMOS2025 contours, with masses lower than $10^{10}$\msun. Only a few galaxies are found outside of the $3\sigma$ contour of the field SFGs. 

For the Spiderweb protocluster passive galaxies, the MSR increases more slowly ($m=0.10$) with stellar mass, as compared to the field passive galaxies estimated using the COSMOS2025 catalogue ($m=0.43$). The same is true, if compared to the MSR from \citet{vanderwel2014}, which is based on HST data for which the relation is even steeper (see also \citealt{Nedkova2021MNRAS.506..928N}). We can see that all of the Spiderweb passive galaxies fall within the COSMOS2025 contours, while simultaneously lying above the field MSR for passive galaxies, thus being larger than the field passive galaxies (see Sec. \ref{sec:intercept} for more details). A similar picture is obtained when comparing to \cite[][see Fig. \ref{fig:msr_discussion}]{Martorano2024ApJ...972..134M}. The passive, X-ray detected galaxies \citep{Tozzi2022A&A...662A..54T} are distributed around our MSR evenly. This suggests that, at fixed stellar mass, X-ray activity does not introduce a systematic size offset relative to the overall passive population. It is also in the line of findings by \citet{Naufal2024ApJ...977...58N}, who showed that $\sim50\%$ of massive quiescent galaxies may host an active galactic nuclei. 

\subsubsection{Mass--size relation intercept evolution}
\label{sec:intercept}

\begin{figure}[ht!]
\includegraphics[width=0.5\textwidth,clip]{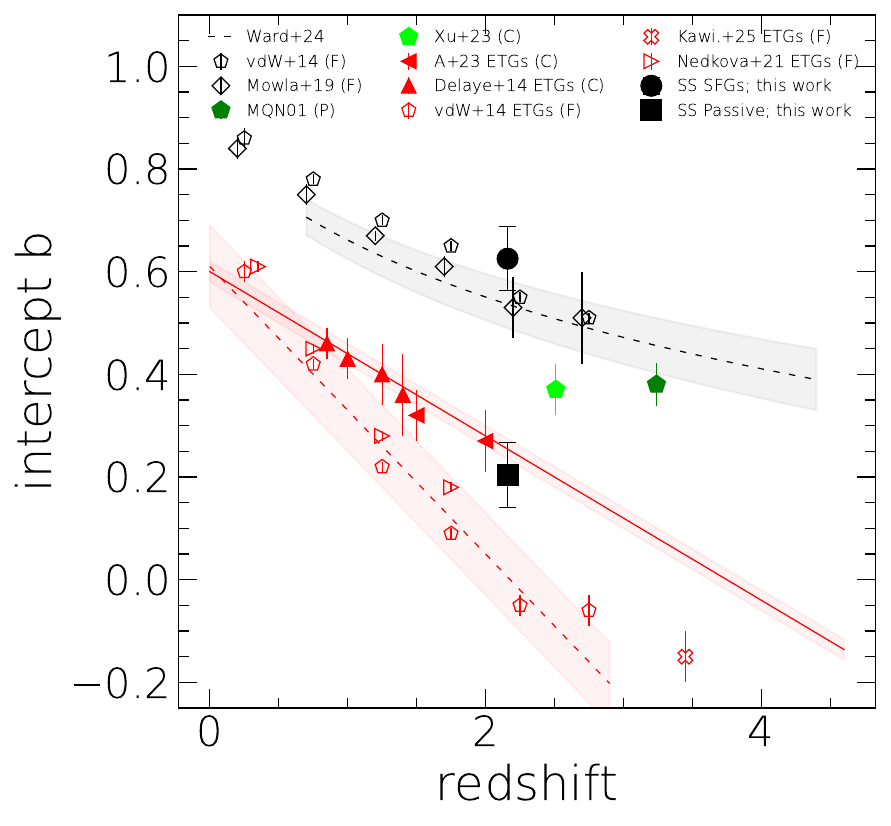}
    \caption{Best-fit MSR intercept as a function of redshift, evaluated at \mstar$=5\times10^{10}$\msun. Black-filled circles and squares show the intercepts of the F182M SS model ($\sim5800\,\mathrm{\AA}$ rest-frame) for SFGs and passive galaxies from this work. Our error bars represent 68\% confidence intervals based on 1\,000 bootstrap resampling. Data from \citet{vanderwel2014,Mowla2019ApJ...880...57M,Nedkova2021MNRAS.506..928N,Ward2024ApJ...962..176W,Kawinwanichakij2025} are shown for comparison with the field LTGs and ETGs. Data from \citet{Xu2023ApJ...951L..21X} and from \citet{Galbiati2025A&A...696A..95G} for MQN01 show SFGs in clusters at $z\sim2.5$ and a protocluster at $z\sim3.2$, respectively (MQN01 intercept from private comm. with W. Wang). Data from \citet{Afanasiev2023A&A...670A..95A} and \citet{Delaye2014MNRAS.441..203D} for ETGs in clusters are shown as red triangles together with the relation from \citet{Afanasiev2023A&A...670A..95A}. P, C, and F, given in the legend, denote protocluster, cluster, and field for clarity. The passive Spiderweb protocluster galaxies intercept is between the field and the cluster intercepts. 
    \label{fig:intercept_z}}
\end{figure}

As shown in several works \citep{vanderwel2014,Delaye2014MNRAS.441..203D,Mowla2019ApJ...880...57M,Nadolny2021A&A...647A..89N,Ward2024ApJ...962..176W,Kawinwanichakij2025}, the typical size of galaxies evolves with redshift. Galaxies at higher redshifts are more compact, with sizes increasing toward lower redshifts. In general terms, SFGs are larger and exhibit slower evolution (the intercept evolves more slowly), whereas passive galaxies are more compact and the MSR intercept evolves more rapidly. There is also an environmental effect at play. \citet{Afanasiev2023A&A...670A..95A} showed that the passive early-type galaxies in clusters are on average $0.2$ -- $0.3$ dex larger than the field ETGs at similar redshifts, while \citet{Kawinwanichakij2025} suggests that the post-merger galaxies at denser environments have a higher bulge-to-total (BT) ratio, while a similar BT ratio in intermediate densities is reached via other mechanisms like violent disk instabilities. 

In Figure \ref{fig:intercept_z}, we show the intercepts of the MSRs from Section \ref{sec:results_MSR} for SFGs and passive galaxies (black filled circle and square) found in the Spiderweb protocluster. Our result for SFGs shows a slightly larger (however, within the uncertainties) intercept $b=0.63\pm0.09$ ($4.26\pm1.2$ kpc) than the field SFGs as estimated using CEERS/JWST data from \citet{Ward2024ApJ...962..176W}, or as given in previous HST-based works \citep{vanderwel2014,Mowla2019ApJ...880...57M}. We also show intercepts corresponding to high-$z$ cluster SFGs \citep{Xu2023ApJ...951L..21X} at $z\sim2.51$, and a protocluster MQN01  \citep{Pensabene2024A&A.MNQ01,Wang2025NatAs} at $z\sim3.2$. Both show lower intercepts for galaxy members (by $\sim0.2$ dex) than for field galaxies and the Spiderweb protocluster. In the case of the J1001 cluster studied in \citet{Xu2023ApJ...951L..21X}, the difference in the evolutionary stage of the overdensity (which has already virialised as a cluster) may play a role in the morphological differences. For the MNQ01 protocluster at $z\sim3.2$, the intercept is slightly lower than that of field galaxies; however, within uncertainties \citep{Ward2024ApJ...962..176W}.

The intercept $b = 0.20\pm0.06$ (or typical size of $1.58\pm0.17$ kpc) of passive Spiderweb protocluster galaxies is found to be $\sim0.2$ dex above the field intercepts from COSMOS2025 catalogue (estimated in this work), \citet{Kawinwanichakij2025}, and \citet{vanderwel2014}; and slightly below ($\sim0.05$ dex; within the uncertainties) the cluster passive galaxies from \citet{Delaye2014MNRAS.441..203D,Afanasiev2023A&A...670A..95A}. This offset suggests that the passive population in Spiderweb protocluster may already include newly quenched galaxies with larger sizes than coeval field passives \citep{Papovich2012ApJ...750...93P,Andreon2018A&A...617A..53A}, rather than requiring only strong size growth of the same passive systems as the protocluster matures.  

If considering SFGs and passive galaxies with \zspec\ confirmation, our conclusions remain the same (as already shown in Sec. \ref{sec:results_SS_MSR}).

\subsection{MSR: Bulge-disk decomposition}
\label{sec:size_with_wave}

\begin{table}
\centering
\begin{tabular}{llccc}
\toprule
Component & Filter & $m$ & $b$ & $\sigma$ \\
\midrule
\multicolumn{5}{c}{\textbf{Star-forming galaxies}} \\
\midrule
\multirow{3}{*}{Single Sérsic} & F115W & $0.36 \pm 0.09$ & $0.66 \pm 0.07$ & $0.24$ \\
 & F182M & $0.33 \pm 0.08$ & $0.63 \pm 0.07$ & $0.22$ \\
 & F410M & $0.22 \pm 0.05$ & $0.48 \pm 0.05$ & $0.17$ \\
\addlinespace
\multirow{3}{*}{Bulge} & F115W & $0.28 \pm 0.12$ & $0.34 \pm 0.11$ & $0.30$ \\
 & F182M & $0.25 \pm 0.12$ & $0.33 \pm 0.11$ & $0.27$ \\
 & F410M & $0.18 \pm 0.12$ & $0.25 \pm 0.11$ & $0.25$ \\
\addlinespace
\multirow{3}{*}{Disk} & F115W & $0.23 \pm 0.12$ & $0.78 \pm 0.07$ & $0.40$ \\
 & F182M & $0.19 \pm 0.12$ & $0.75 \pm 0.07$ & $0.39$ \\
 & F410M & $0.06 \pm 0.10$ & $0.61 \pm 0.06$ & $0.37$ \\
\addlinespace
\midrule
\multicolumn{5}{c}{\textbf{Passive galaxies}} \\
\midrule
\multirow{3}{*}{Single Sérsic} & F115W & $0.07 \pm 0.17$ & $0.22 \pm 0.07$ & $0.23$ \\
 & F182M & $0.10 \pm 0.16$ & $0.20 \pm 0.06$ & $0.22$ \\
 & F410M & $0.23 \pm 0.12$ & $0.15 \pm 0.05$ & $0.18$ \\
\addlinespace
\multirow{3}{*}{Bulge} & F115W & $0.17 \pm 0.19$ & $-0.17 \pm 0.08$ & $0.29$ \\
 & F182M & $0.20 \pm 0.15$ & $-0.14 \pm 0.07$ & $0.26$ \\
 & F410M & $0.25 \pm 0.13$ & $-0.09 \pm 0.06$ & $0.22$ \\
\addlinespace
\multirow{3}{*}{Disk} & F115W & $0.11 \pm 0.14$ & $0.44 \pm 0.05$ & $0.19$ \\
 & F182M & $0.12 \pm 0.13$ & $0.44 \pm 0.05$ & $0.18$ \\
 & F410M & $0.12 \pm 0.17$ & $0.44 \pm 0.05$ & $0.20$ \\
\addlinespace

\bottomrule
\end{tabular}
\caption{\label{tab:MSR_results} Mass--size relation fits. Linear fits follow
$\log(R_{\rm eff}/{\rm kpc}) = m\,(\log M_\star - \log(5\times10^{10}M_\odot)) + b$.
The upper block shows star-forming galaxies and the lower block passive galaxies.
$\sigma$ indicates the scatter of galaxies around the best-fit relation.
Uncertainties correspond to bootstrap errors.}
\end{table}

\begin{figure*}[ht!]
\includegraphics[width=\textwidth,clip]{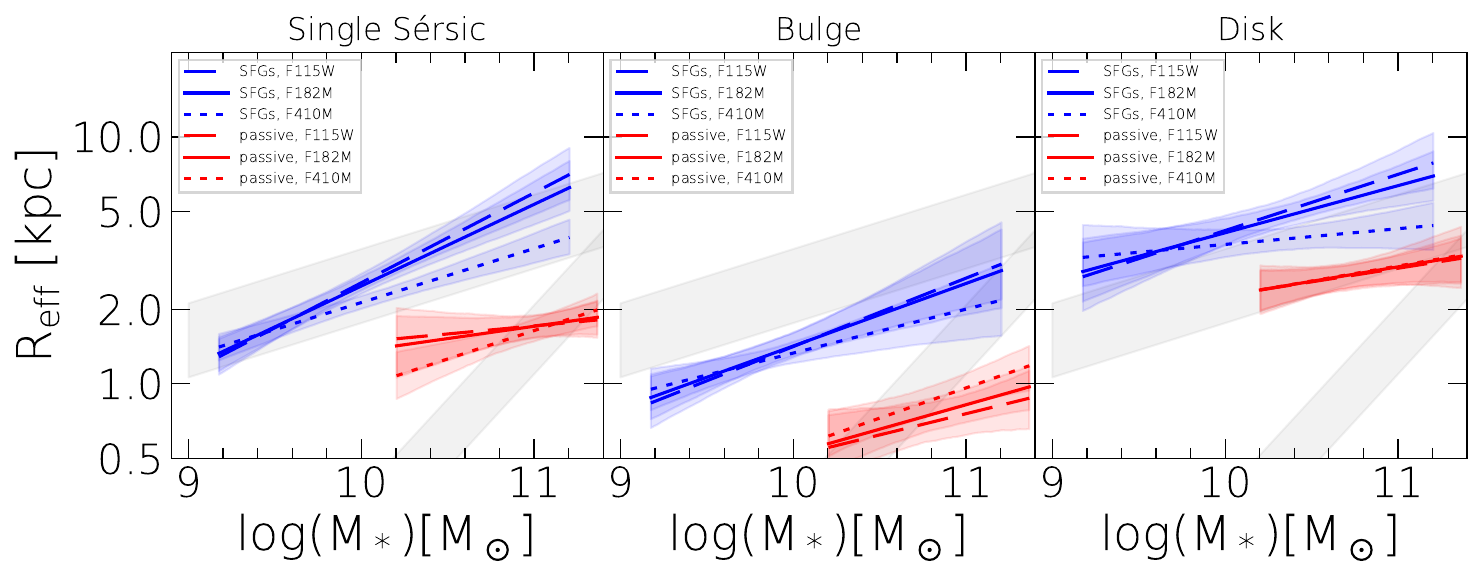}
    \caption{Mass--size relation for SFGs. Best fits are shown for Single S\'ersic, bulge and disk components in different filters. The MSR from \citet{vanderwel2014} is shown for comparison. Bulge components are always more compact than disk components. The best-fit parameters are given in Table \ref{tab:MSR_results}.
    \label{fig:msr_bd_filter}}
\end{figure*}

\begin{figure*}[ht!]
\includegraphics[width=\textwidth,clip]{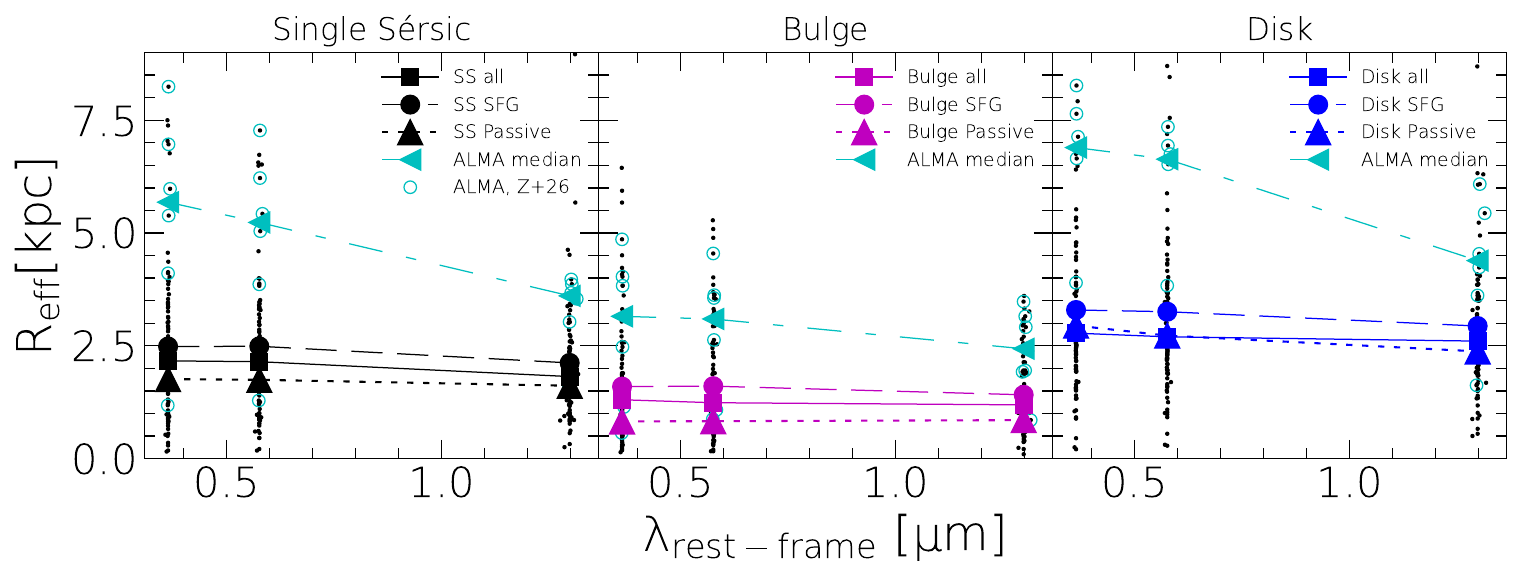}
    \caption{Effective radius as a function of rest-frame wavelength. Black small points show the individual galaxies from our morphological sample. Large squares, circles, and triangles in black, magenta and blue show the median values for all galaxies, SFGs and passive galaxies measured for a single S\'ersic model, bulge and disk components. Cyan left-triangles show ALMA-detected subsample from \citet{Zhang2025A&A}.  
    \label{fig:reff_wave}}
\end{figure*}

In this section, we will examine the size evolution of our morphological sample of SFGs and passive galaxies in the context of the BD decomposition (the same as shown in Fig. \ref{fig:msr}), and in comparison with our previous SS results. Furthermore, we will show the size dependency with wavelength. 

In Figure \ref{fig:msr_bd_filter}, we show the results of the fitting of the MSR for SFGs and passive galaxies using all the filters per component (SS, bulge, disk). The general trend is that for all the components, the MSR for SFGs measured in F115W and F182M have essentially the same slope, while for F410M the MSRs are the flattest, within the uncertainties. The measured MSR intercepts for both the disc and bulge components in the F115W and F182M bands are consistent, albeit those values are higher than those of F410M, however still within the uncertainties. It is also clear that the MSR from SS model (regardless of the filter) is broadly consistent with the MSR from \citet{vanderwel2014}. Considering the bulge and disk components are below and above the SS MSR, but within the scatter of the \citet{vanderwel2014} MSR. The best-fit parameters and scatter for each fit can be found in Table \ref{tab:MSR_results}. 

The picture for passive galaxies shows similar general trends: the bulge component is more compact, while the disk component is larger, with the MSR based on the SS model being in between. The difference between SFGs and passive galaxies is that the latter show less variation from one filter to another for each component. This indicates differences in the internal composition of the two subsamples of SFG and passive galaxies. 

To illustrate this in another manner, we show the effective radii $R_{\rm eff}$ as a function of rest-frame wavelength in Figure \ref{fig:reff_wave}, separately for each component. Considering the SS model (left panel), we can see that, on average, for our morphological sample of Spiderweb protocluster galaxies (black squares), the median sizes remain fairly constant, being slightly more compact at NIR rest-frame wavelengths (as found in recent JWST studies at $z>1.5$, e.g., \citealt{Cheng2024ApJ...977..165J,Ji2024arXiv,Allen2025}). When we separate our sample into SFGs (filled circles) and passive (filled triangles) galaxies, the former are slightly larger, while the latter are slightly smaller than the general population, with similar variation with rest-frame wavelength. 

A similar pattern can be seen for the bulge and disk components, with bulges being the most compact ones. A very mild evolution of the size with rest-frame wavelength is observed for these components, independent of the subsamples. 

In Figure \ref{fig:reff_wave} we also include ALMA-detected galaxies studied in \citet{Zhang2025A&A}. They found that these ALMA-detected galaxies display larger stellar components as compared to the general population of Spiderweb SFGs, with their sizes decreasing strongly with rest-frame wavelength. Indeed, the effective radii for these objects decrease drastically from F115W through F182M to F410M bands. This trend is also strongly visible for the disk component, while the bulge component being above other subsamples show mild size evolution with wavelength. Thus, we not only confirm the finding by \citet{Zhang2025A&A} using different methods, but here we extend it by showing that the disk component of the stellar body of ALMA-detected galaxies show similar trend.

\subsection{Passive--density relation}
\label{sec:results_environ}
\begin{figure}[ht]
\includegraphics[width=0.5\textwidth,clip]{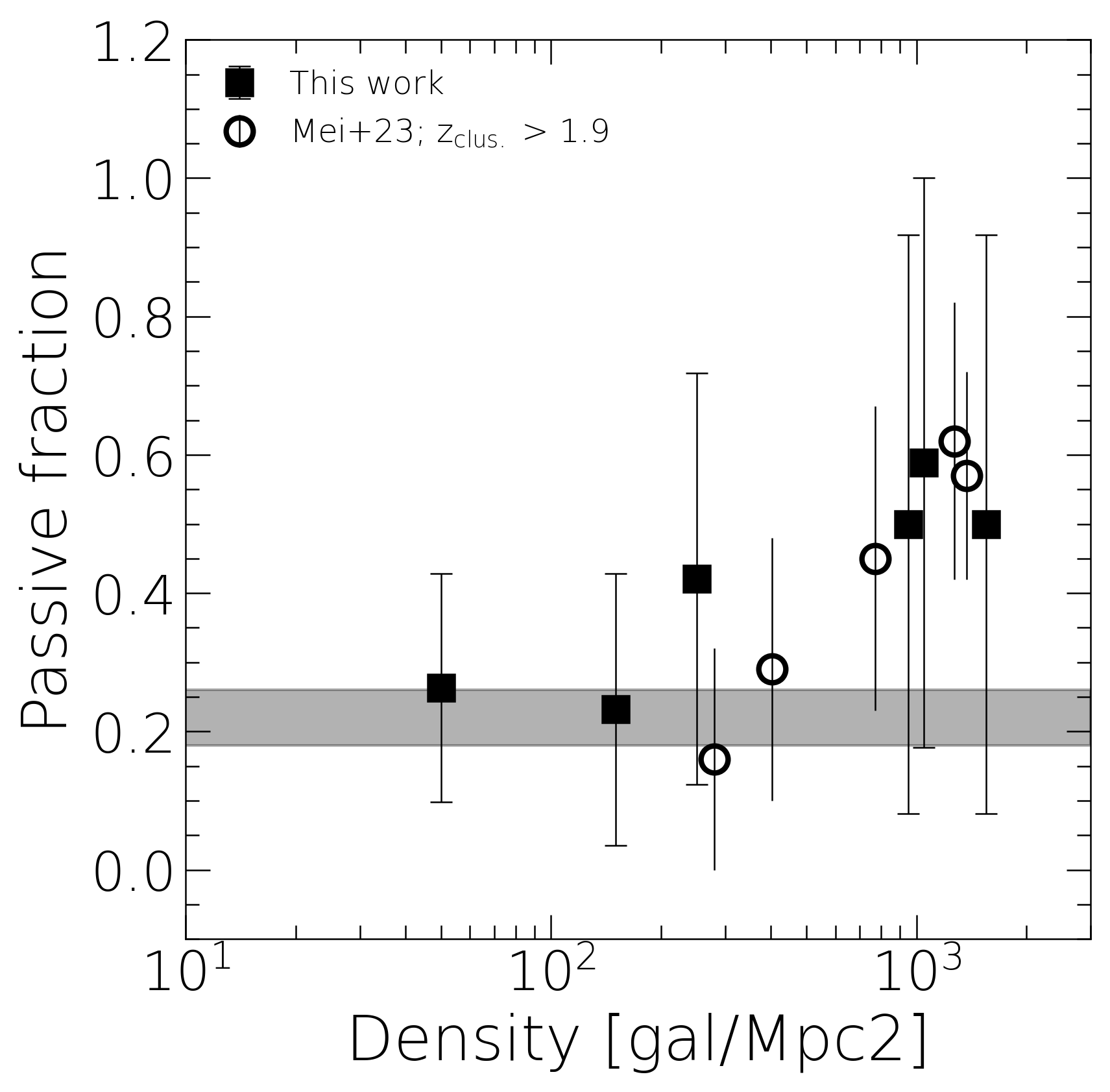}
    \caption{Fraction of passive galaxies as a function of projected local density. Error bars represent 68\% confidence intervals estimated using the formalism of \citet{Wilson01061927}. Data from \citet{Mei2023A&A...670A..58M} for clusters are shown for comparison. Grey horizontal shaded region shows field passive fraction \citep{yang2025_QGdensities}.
    \label{fig:sigma_passiveFraction}}
\end{figure}

Figure \ref{fig:sigma_passiveFraction} shows the fraction of passive (0.3 dex below the SFMS, see Sec. \ref{sec:Sample}) galaxies as a function of increasing local density ($\Sigma_3$, Sec. \ref{sec:methods_environment}). We find that the passive fraction increases with density from $\sim20\%$ at densities of $\sim50\,\rm [gal/Mpc^2]$ (consistent with the field fraction) to about $60\%$ at $\sim1100\,\rm [gal/Mpc^2]$. This is consistent with results from the CLARA survey \citep{Wylezalek2013ApJ...769...79W} at $z$ between 1.9 and 2.8 from \citet{Mei2023A&A...670A..58M}, shown in the same Figure. Because \citet{Mei2023A&A...670A..58M} use a UVJ-based passive selection while we define passive galaxies as lying below the SFMS, this agreement should be interpreted as consistency in the environmental trend, rather than as a one-to-one comparison of galaxy classes. Compared to the field passive galaxies (grey shaded region), we can see that at densities above 300 galaxies per Mpc$^2$, the passive fraction began to depart toward higher fractions.

{  
Given that this increase in passive fraction is measured using a member catalogue assembled from several follow-up surveys with different spatial coverages, we tested whether the observed passive-density relation could be affected by the non-uniform sampling of the protocluster field (Appendix~\ref{sec_app:footprint}). To this end, we repeated the analysis after restricting the sample to galaxies located within the central HST/WFC3 grism footprint, shown in Fig.~\ref{fig:footprint}. This region corresponds to the area studied by \citet{Naufal2024ApJ...977...58N}, where passive members of the Spiderweb protocluster were spectroscopically confirmed, and is also covered by the main ancillary datasets used in this work. Using this restricted sample, we recover a passive--density relation consistent with that obtained from the full catalogue, with only minor changes in the individual density bins (Fig. \ref{fig:sigma3_hst_grism}). This test indicates that the environmental trend shown in Fig.~\ref{fig:sigma_passiveFraction} is not driven by the uneven spatial coverage of the contributing surveys.
}

To test if the global environment influences the passive fraction, we estimated the passive fraction as a function of distance from the centre normalised to R$_{200}$ up to $\sim2\times$R$_{200}$ due to the JWST coverage (see Fig. \ref{fig:ra_dec}). We show the result in Figure \ref{fig:RR200_passiveFraction}, Appendix \ref{sec_app:environment}. The three bins shown, although consistent with the field fraction within the uncertainties, present a systematically higher passive fraction of $\sim30$--$35\%$. Thus, it suggests that the global environment does not have an impact on the passive fraction, at least up to $\sim1.25\times$R$_{200}$ (where our last bin is located).

We also tested if the environment has an impact on the \sigmasfr, S\'ersic index $n$ or size $R_{\rm eff}$. We find no correlation between $\Sigma_3$ and  effective radius $R_{\rm eff}$ (with Spearman's $\rho=0.01$), or \sigmasfr\ ($\rho=-0.06$), with low statistical significance ($p=0.97$ and $0.63$, respectively). Regarding the S\'ersic indices in all the filters, we find a weak positive correlation with $\sim \rho=0.30$, however, only in the case of F410M this correlation is statistically significant ($p=0.009$). F115W and F182M have $p\sim0.03$, thus indicating a moderate monotonic relationship (see Fig. \ref{fig_app_sigma3_n}). {  Finally, testing more conservative passive selection (as described in Sec. \ref{sec:results_SS_MSR}), we find similar results that do not significantly change our conclusions.}

\section{Discussion}
\label{sec:discussion}
In this work, we present morphological analysis of all the known Spiderweb protocluster member galaxies found in the JWST FoV (114 in total). We examine the mass--size relation and the redshift evolution of the characteristic size (intercept), placing our results in the context of coeval field and cluster populations. We also investigate the influence of the environment on passive fraction. In what follows, we will discuss our results in the broader context of galaxy evolution.

\subsection{Mass--size relation}
\label{sec:discussion_MSR}
\begin{figure}[ht!]
\includegraphics[width=0.5\textwidth,clip]{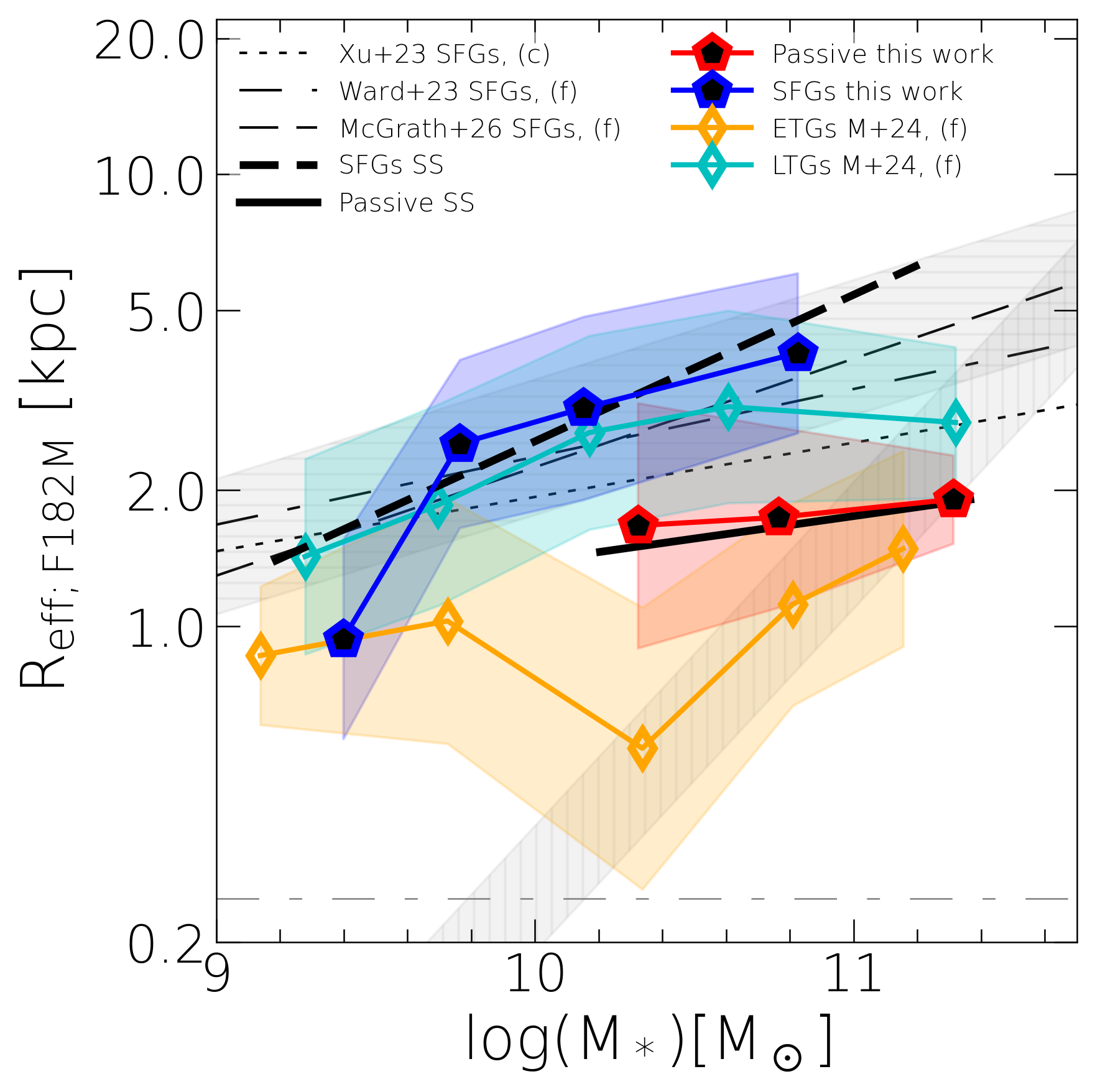}
    \caption{Mass--size relation. Blue and red pentagons show median values of SFGs and passive galaxies from this work. Cyan and orange diamonds show median values for field galaxies from \citet{Martorano2024ApJ...972..134M} catalogue selected as described in Section \ref{sec:comparision_data}. Grey shaded horizontally and vertically hatched regions show LTGs and ETGs MSR from \citet{vanderwel2014}. Dotted black line is the MSR for SFG from cluster studied by \citet{Xu2023ApJ...951L..21X}, while the dot-dashed and dashed lines are the MSR for the field SFGs from \citet{Ward2024ApJ...962..176W} and \citet{McGrath2026ApJ...999L...6M}. 
    \label{fig:msr_discussion}}
\end{figure}

For several decades, it has been systematically shown that the most massive galaxies are also the largest. This constitutes the mass--size relation \citep{vanderwel2014,Mowla2019ApJ...880...57M,Nadolny2021A&A...647A..89N,Nedkova2021MNRAS.506..928N,Nedkova2024MNRAS.532.3747N,Martorano2024ApJ...972..134M,QuilleyEuclidMorpho2025}. At fixed stellar mass, galaxy sizes decrease with redshift, and this evolution is faster for passive (or ETGs) than for SFGs (or LTGs) \citep[e.g.][]{Ward2024ApJ...962..176W}. In the context of a cluster environment, \citet{Afanasiev2023A&A...670A..95A} shows that ETGs in dense cluster environments grow faster at higher redshift and then evolve more slowly than field ETGs. 

In Figure \ref{fig:msr_discussion} we compare median values (with $1\sigma$ deviation) for our SFGs and passive galaxies with the field population from \citet{Martorano2024ApJ...972..134M} in stellar-mass bins (from \mstar$=10^9$\msun, with 0.5 dex width; see Sec. \ref{sec:comparision_data}). For SFGs, the \citet{Martorano2024ApJ...972..134M} UVJ-selected medians (cyan diamonds) follow our fit up to \mstar$\sim10^{10.6}$\msun; in our highest-mass bin (\mstar$>10^{10.5}$\msun, six objects, only one above \mstar$>10^{11}$\msun), both data sets show a mild downturn. This behaviour at high-mass end is broadly consistent with \citet{Ward2024ApJ...962..176W}, who find a nearly redshift-invariant SFG MSR slope with evolving intercept, and with \citet{Martorano2024ApJ...972..134M}, who report flatter near-IR SFG relations and stronger size offsets at high mass. It is also in line with \citet{McGrath2026ApJ...999L...6M}, who show that late-type galaxies have increasingly strong mass-dependent color gradients, which can enhance apparent compactness in the most massive systems. In dense environments, \citet{Xu2023ApJ...951L..21X} find that massive cluster SFGs at $z\sim2.5$ are systematically more compact than field counterparts, suggesting accelerated structural evolution; our most massive Spiderweb SFGs may follow the same trend, but the current sample size does not allow a firm conclusion.

Our passive galaxies are slightly larger as compared to the field UVJ-selected passive galaxies from \citet[][within $1\sigma$ deviation]{Martorano2024ApJ...972..134M}, in particular for the lower-mass bin (the result does not change with the number of bins). This makes our fit flatter with a larger typical size of the passive galaxy population in the protocluster.
On the other hand, the median values from \citet{Martorano2024ApJ...972..134M} follow the fit from \citet{vanderwel2014} for passive galaxies down to \mstar$\sim10^{10.4}$\msun, while in the two lower mass bins their median $R_{\rm eff}$ values increase considerably. At lower redshift ($z\sim1.75$) \citet{Nedkova2021MNRAS.506..928N} found that a double power law describes the passive MSR better with a flattening of this relation at lower stellar masses. The study by \citet{Afanasiev2023A&A...670A..95A} revealed similar behaviour comparing clusters to the field, where cluster passive galaxies were consistently larger than the field. This will be better illustrated in the next section.


\subsection{MSR intercept evolution}
\label{sec:discussion_intercept}
The MSR intercept has been shown to evolve with redshift, with different slopes for SFGs (or LTGs) and passive (or ETGs) \citep{vanderwel2014,Mowla2019ApJ...880...57M,Nedkova2021MNRAS.506..928N,Afanasiev2023A&A...670A..95A,Ward2024ApJ...962..176W}. The intercept evolution for SFGs is slower, while passive galaxies exhibit a much faster (steeper) increase in intercept from the earlier universe to today. Our SFGs show a consistent intercept at a given redshift with the field SFGs \citep[see Fig. \ref{fig:intercept_z}]{Ward2024ApJ...962..176W}. The intercept of passive galaxies found in the Spiderweb protocluster falls above the field and just below (within the uncertainties) the clusters' environments (at similar redshift, $z\sim2$), the latter being larger.

{  Several caveats should be kept in mind when interpreting this comparison. Although our F182M measurements probe the rest-frame optical emission ($\sim5800$,\AA) and are therefore broadly comparable to the HST- and JWST-based literature relations used here, the reference samples were derived with different filters, depths, PSFs, profile-fitting methods, and passive-galaxy selections. In particular, some comparison works use UVJ- or NUV--$r$--$J$-selected passive galaxies, whereas our passive subsample is defined relative to the SFMS. Such differences can introduce systematic shifts in the fitted slopes and intercepts. Indeed, previous studies have shown that different quiescent-galaxy definitions, including selections based on sSFR, UVJ colours, and structural parameters, generally recover the same broad MSR trends, but do not necessarily select identical galaxy populations \citep{Wu2018_FastSlowQuiescence,Nedkova2021MNRAS.506..928N}. This is particularly relevant for transition or recently quenched systems: \citet{Fang2018_UVJ_CANDELS} showed that galaxies lying below the SFMS can occupy an intermediate region of UVJ space and have smaller radii than main-sequence galaxies, while \citet{Lustig2023_MassiveQuiescent_z3} emphasised that UVJ-selected high-redshift quiescent samples may also be affected by contamination from dusty star-forming galaxies and incompleteness for young quiescent systems.

We repeat our analysis by applying a stricter passive-galaxy selection, now defined as objects lying 1 dex below the SFMS. This test removes four galaxies from the passive subsample and leaves our results qualitatively unchanged. We therefore interpret the location of the Spiderweb protocluster passive MSR between the field and cluster relations as evidence for an intermediate structural regime, rather than as a precise quantitative evolutionary track.}

While it was observed that cluster and field passive galaxies display a defined evolutionary path \citep{Delaye2014MNRAS.441..203D,Afanasiev2023A&A...670A..95A}, this work indicates that the protocluster environment may be in a different evolutionary stage. At lower redshifts ($z < 1$), the field and cluster passive galaxy MSR intercepts evolve in a similar way \citep{Figueira2024A&A...687A.117F}. At higher redshifts ($z > 1.5$), however, the differences between the two populations become significant, cluster galaxies being larger (higher intercept). Our passive protocluster sample, which is found in a different environment from the field, but not yet fully virialised as a cluster, is found significantly above the field passive galaxies, and just below, within the uncertainties, the cluster passive galaxies. While the passive--density relation is already observed to be in place (see Fig. \ref{fig:sigma_passiveFraction}), this offset does not necessarily imply that the same passive ETGs in the protocluster have already undergone strong in-situ size growth. A plausible alternative is that the passive population is being built up through a progenitor bias effect, in which newly quenched galaxies enter the passive sample at larger sizes than coeval field passive galaxies. In the field, many high-redshift quiescent galaxies are thought to pass through a compact phase produced by dissipative compaction before quenching, yielding compact newly passive systems at $z\gtrsim2$ \citep{Barro2013ApJ...765..104B,Dekel2014MNRAS.438.1870D,Barro2017ApJ...840...47B}. In a maturing protocluster, however, environmental quenching and preprocessing may instead add already larger galaxies to the passive population through strangulation, harassment, ram-pressure stripping, or related processes \citep{Strazzullo2013ApJ...772..118S,Delaye2014MNRAS.441..203D}. This interpretation is consistent with studies showing that dense environments can host larger quiescent galaxies, or equivalently, a deficit of the most compact systems, relative to the field at fixed stellar mass \citep{Papovich2012ApJ...750...93P,Andreon2018A&A...617A..53A}. Finally, some additional size growth through mergers is still expected during the assembly of the densest structures \citep{Matharu2019MNRAS.484..595M,DD_Shi2024ApJ...963...21S}

\subsection{Bulge--disk decomposition}

Our bulge--disk decomposition reveals that, for both SFGs and passive galaxies, the bulge component is more compact (by $\sim0.3$--$0.4$ dex) than the corresponding disc component in all filters, as indicated by the intercept $b$ (or typical size evaluated at \mstar$=10^{10.5}$\msun; see Table \ref{tab:MSR_results}). For SFGs, both components have similar slopes and intercepts in the F115W and F182M bands, and run parallel to the SFG MSR from \citet{vanderwel2014}. The bulge and disc components of SFGs in F410M show relatively flatter and more compact MSRs compared to F115W and F182M. For passive galaxies, bulge and disc components show very similar fitting results in all filters, being consistent with objects that resemble much simpler internal morphology (also indicated by a higher S\'ersic index; see Fig. \ref{fig:ss_index}).

Considering the size evolution with the wavelength shown in Figure \ref{fig:reff_wave}, we find that independently of the model, SFGs are always larger than passive galaxies. Regarding the bulge component, it is larger in SFGs as compared to passive galaxies by $\sim$160\%, 130\%, and 50\% in F115W, F182M, and F410M respectively. The study by  \citet{Dimauro2019MNRAS.489.4135D}, reveals that bulges in field SFGs are $>20\%$ larger than bulges in passive objects at $z\sim1.75$. The difference between bulges in SFGs and passive galaxies in the Spiderweb protocluster is larger than the difference between the bulges of field SFGs and passive galaxies. This difference may be attributed to the selection process of galaxies for the BD decomposition, but also to the redshift range studied. This difference in redshift between $z=2.16$ and $z=1.75$ (the last bin in \citealt{Dimauro2019MNRAS.489.4135D}) is the time of a steep size evolution of the field passive galaxies (red dotted line in Fig. \ref{fig:intercept_z}). Finally, we can observe that bulge sizes decrease with increasing wavelength. 

The difference between SFGs and passive galaxies' disk component is much smaller comparing to the bulge component. We found the differences of $<20\%$ in all the bands. This is consistent with results for disks in field SFGs and passive galaxies at $z\sim1.75$ from \citet{Dimauro2019MNRAS.489.4135D}. 

As described by \citet{Zhang2025A&A}, the ALMA-detected dusty SFGs appear more evolved than coeval SMGs and show a steep decrease of $R_{\rm eff}$ toward redder wavelengths (Fig. \ref{fig:reff_wave}). We confirm this behaviour and show that it is present in both bulge and disk components, with a stronger gradient in the disk. A consistent interpretation is that these galaxies combine unusually extended stellar disks with centrally concentrated, dust-obscured star formation: bluer rest-frame light traces the outer disk, while redder wavelengths increasingly sample compact central regions. This naturally explains why their bulges lie closer to the median Spiderweb trend, whereas their disks remain above the general population, in line with the accelerated-growth scenario in dense, likely infalling substructures.


\subsection{Environment and passive--density relation}
\label{sec:discussion_MDR}

The influence of the environment on galaxy evolution has been studied at a wide range of redshifts and using different methods to measure the environment. In particular, the local projected density has been shown to correlate with several key galaxies' parameters, like the fraction of ETGs and passive galaxies. Local galaxy clusters ($z\sim0$) show a high ETG fraction at virtually all local densities \citep{Dressler1980ApJ...236..351D,Dressler1997ApJ...490..577D}, varying from 40--50\% up to 90\% in the highest local densities. At an intermediate redshift ($0.7< z < 1$), ETG fractions in clusters were observed to be slightly lower, however, showing a steady increase of the ETG with the local density from $\sim 40\%$ up to $80\%$ at densities of $\sim1000$ galaxies per Mpc$^2$ \citep{Postman2005ApJ,Simard2009A&A...508.1141S}. 
At higher redshifts ($1.3 < z < 2.8$) \citep{Mei2023A&A...670A..58M} find that the passive fraction increases significantly above densities of $300-400$ galaxies per Mpc$^2$, establishing that passive-- and morphology--density relations are in place at $z\sim2$.
These results suggest that at higher redshift and lower densities ($<300-400$ galaxies per Mpc$^2$), galaxies did not have enough time to go through the morphological transformation (either via mergers, harassment, or morphological quenching, to name a few). At high redshift and high densities ($>400$ galaxies per Mpc$^2$) we can see a steep increase of the passive fraction with density reaching $\sim60\%$ at the highest densities ($>1000$ galaxies per Mpc$^2$; this work and \citealt{Mei2023A&A...670A..58M}). 

Our passive fraction in projected local density (Fig. \ref{fig:sigma_passiveFraction}), is consistent with the results from \citet{Mei2023A&A...670A..58M}. While the low number counts in most of the bins (between six and one sources) give relatively larger uncertainties, our measurements show a substantial increase in the passive fraction from being consistent with the field fraction of $\sim20\%$, to increase as high as $\sim60\%$ at densities of $\sim1000$ galaxies per Mpc$^2$. Using only spectroscopically confirmed Spiderweb member galaxies does not change our conclusions. 

To test if the passive fraction depend on the global protocluster environment, we estimate the density of galaxies around the centre of the Spiderweb protocluster, the radio galaxy PSK1138-262 (Fig. \ref{fig:RR200_passiveFraction}), up to $2\times{\rm R_{200}}$. While we do not see a correlation, it is clear (within the uncertainties) that the fraction of passive galaxies is systematically higher (by $\sim 10\%$) than the field passive galaxies. This suggests that the global environment plays at most a secondary role, setting an overall offset above the field, whereas the strong increase in passive fraction is primarily driven by the local density traced by $\Sigma_3$. 

Using the density measurements and F410M S\'ersic index measurement, we find a weak, however statistically significant, correlation between both values ($\rho=0.33$, $p = 0.009$; Fig. \ref{fig_app_sigma3_n}). Weaker and statistically significant correlation is found for F115W  ($\rho=0.27$, $p = 0.03$) and F182M  ($\rho=0.30$, $p = 0.02$). This indicates that galaxies with more concentrated light distribution (higher $n$) at longer wavelengths (redder rest-frame colours) are found in a denser environment. This reflects the morphology--density relation, with increasing passive fraction at higher (local) densities.

Finally, we found no correlation between the local density and effective radii and \sigmasfr\ (Fig. \ref{fig_app_sigma3_n}). While we can see the correlation of the light concentration (as measured by S\'ersic index), the sizes of galaxies seem to be independent of local environment. The same is true for \sigmasfr, what is expected since the size (as shown above) and SFR (as shown in \citealt{Pérez-Martínez2023MNRAS.518.1707P}) do not depend on the local density peaks.


\section{Conclusions}
In this work, we present a homogeneous morphological analysis of the Spiderweb protocluster at $z=2.16$ using JWST/NIRCam imaging. Our main results are the following:

\begin{itemize}
\item  We provide the first homogeneous structural analysis of the full Spiderweb protocluster member sample within the JWST FoV (up to $\sim2\times R_{200}$), including different member classes (HAEs, LAEs, and PBEs), and derive consistent single-S\'ersic and bulge--disc parameters.

\item  The SFG mass--size relation is broadly consistent with the field relation, with a mildly higher slope and slightly larger sizes at the high-mass end within uncertainties (Fig. \ref{fig:msr}).

\item  Passive galaxies show a flatter MSR than field passive populations. Their intercept (typical size at stellar mass of $5\times10^{10}$\msun) lies above the field and slightly below virialised clusters at similar redshift, indicating an intermediate evolutionary stage in which the passive population is already offset from the field, likely through the addition of larger newly quenched members, but is not yet fully cluster-like (Fig. \ref{fig:intercept_z}).

\item  Bulge--disc decomposition shows that bulges are systematically more compact than discs for both SFGs and passive galaxies in all filters. For SFGs, MSR slopes of both components are steeper in bluer filters, while for passive galaxies slopes and intercepts are more stable across filters and components (Figs. \ref{fig:msr_bd_filter} and \ref{fig:reff_wave}) indicating differences in the internal morphology, the former being more complex. 

\item  Galaxy sizes decrease mildly with increasing wavelength, whereas ALMA-detected dusty SFGs show a much steeper wavelength dependence. This steep trend is also seen in their bulge and disc components, consistent with centrally concentrated obscured star formation.

\item  The passive fraction depends primarily on local density, not on global clustercentric distance. It is field-like at low densities, rises strongly above $\Sigma\sim200$--$300\,{\rm gal\,Mpc^{-2}}$, and reaches $\sim60\%$ in the densest regions ($>1000\,{\rm gal\,Mpc^{-2}}$), demonstrating that the passive--density relation is already established at $z=2.16$ (Figs. \ref{fig:sigma_passiveFraction} and \ref{fig:RR200_passiveFraction}).

\item  In addition to the passive-fraction trend, we find a weak but statistically significant correlation between local density and S\'ersic index (strongest in F410M), while no clear correlation is found between local density and effective radius or $\Sigma_{\rm SFR}$ (Fig. \ref{fig_app_sigma3_n}).
\end{itemize}

Overall, the Spiderweb protocluster appears to be in a transition regime: the MSR intercept of our passive Spiderweb galaxies lies above the field and slightly below the cluster passive galaxy intercepts, suggesting that the passive population is being assembled in an environment that can add larger newly quenched systems, while some further structural evolution may still occur before the system reaches the fully virialised cluster stage. The passive fraction is already comparable to that of clusters at similar redshift and is primarily driven by local density, while the structural evolution of passive galaxies is still ongoing.


\begin{acknowledgements}
J.N. acknowledge the support of the National Science Centre, Poland, through the SONATA BIS grant 2018/30/E/ST9/00208. J.N.~acknowledges the support of the Polish National Agency for Academic Exchange (NAWA) Bekker grant BPN/BEK/2023/1/00271, and the kind hospitality of the IAC.

HD and JMPM acknowledge support from the Agencia Estatal de Investigación del Ministerio de Ciencia, Innovación y Universidades (MCIU/AEI) under grant (Construcción de cúmulos de galaxias en formación a través de la formación estelar oscurecida por el polvo) and the European Regional Development Fund (ERDF) with reference (PID2022-143243NB-I00/DOI:10.13039/501100011033).
MHC  acknowledge financial support from the State Research Agency of the Spanish Ministry of Science and Innovation (AEI-MCINN) under the grant “BASALT”
with reference PID2021-126838NB-I00.
YK acknowledge support from JSPS KAKENHI grant Nos. 22K21349, 23H01219, 24H00002, and
JSPS Core-to-Core Program (JPJSCCA20210003).

KD acknowledges financial support from JSPS KAKENHI Grant Number 25K23411.

TK acknowledges financial support from JSPS KAKENHI Grant Numbers 24H00002 (Specially Promoted Research by T. Kodama et al.), and 22K21349 (International Leading Research by S. Miyazaki et al.).

The authors wish to acknowledge the contribution of the IAC High-Performance, Computing support team and hardware facilities to the results of this research.

\end{acknowledgements}

\bibliographystyle{aa} 
\bibliography{biblography}

\begin{appendix}
\section{SExtractor setup}
\label{sec_app:sextractor}
In Table \ref{tab:sextractor} we show relevant parameters used in Hot and Cold run of SExtractor. See details in Sec. \ref{sec:methods_parametric}.

\begin{table}
	\caption[]{Main configuration parameters used in SExtractor HDR run on NIRCam/F182M}
	\label{tab:sextractor}
	\begin{tabular}{rll}
		\hline\hline 
		Parameter &  hot & cold \\ 
		\hline
		\texttt{DETECTED\_MINAREA} [px] 	& 10     & 20 \\
		\texttt{DETECT\_THRESH} [$\sigma$]	& 1.8   & 4.5  \\
		\texttt{ANALYSIS\_THRESH} [$\sigma$] 	& 1   &5  \\
		\texttt{FILTER\_NAME} 		& \texttt{gauss} & \texttt{gauss}  \\
		\texttt{DEBLEND\_NTHRESH} [branch] 	& 32    & 64  \\
		\texttt{DEBLEND\_MINCONT} [fraction]	& 0.008 & 0.04 \\	
		\texttt{BACK\_SIZE} [px]		& 64   & 64 \\	
		\texttt{BACK\_FILTERSIZE} [px] 	& 3     & 3 \\		
		\texttt{BACKPHOTO\_TYPE} 	& GLOBAL & GLOBAL \\	
		\texttt{BACKPHOTO\_THICK} [px]	& 24    & 24 \\	
		\hline
	\end{tabular}
\end{table}

\section{S\'ersic index}
\begin{figure}[ht!]
\includegraphics[width=0.5\textwidth,clip]{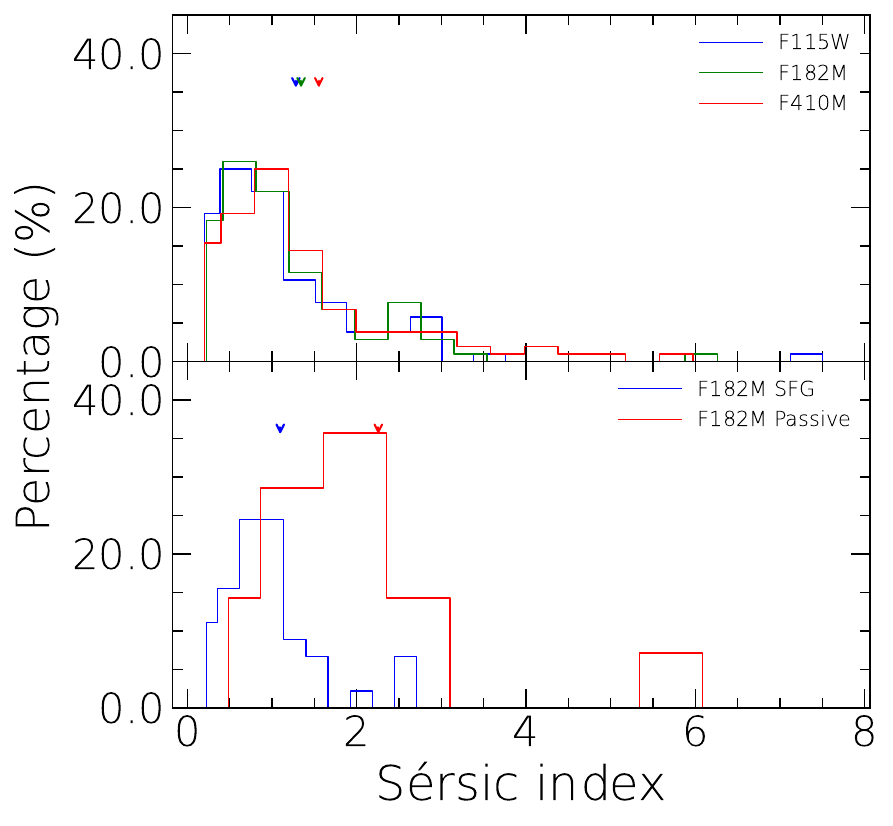}
    \caption{Top: the overall sample S\'ersic index in F115W, F182M and F410M. Bottom: S\'ersic index for SFGs and passive galaxies in F182M filter. Arrows indicate the average value of the S\'ersic index for a given subsample, denoted by the same colour as the histograms.
    \label{fig:ss_index}}
\end{figure}

{  In the Figure \ref{fig:ss_index} we show the distribution of S\'ersic indices for all the galaxies per filter (top panle) and distribution of S\'ersic indices for SFGs and passive galaxies (bottom panel), with the average of 1.1 and 2.25 for SFGs and passive galaxies. 

The bottom panel of Fig. \ref{fig:ss_index} shows that passive galaxies have, on average, higher S\'ersic indices than SFGs, but the two distributions overlap substantially. This overlap indicates that passive galaxies in the Spiderweb protocluster are not uniformly spheroid-dominated systems. Visual inspection of the single-S\'ersic fits, residuals, and NIRCam cutouts confirms that several passive galaxies retain extended or disk-like components, consistent with their relatively low S\'ersic indices ($n<2.5$). Since our passive-galaxy classification is based on star-formation activity rather than morphology, this suggests that quenching may precede the complete morphological transformation of at least part of the passive population.}

\section{Footprint Effects on the Passive–Density Relation}
\label{sec_app:footprint}

\begin{figure}[ht!]
\includegraphics[width=0.5\textwidth,clip]{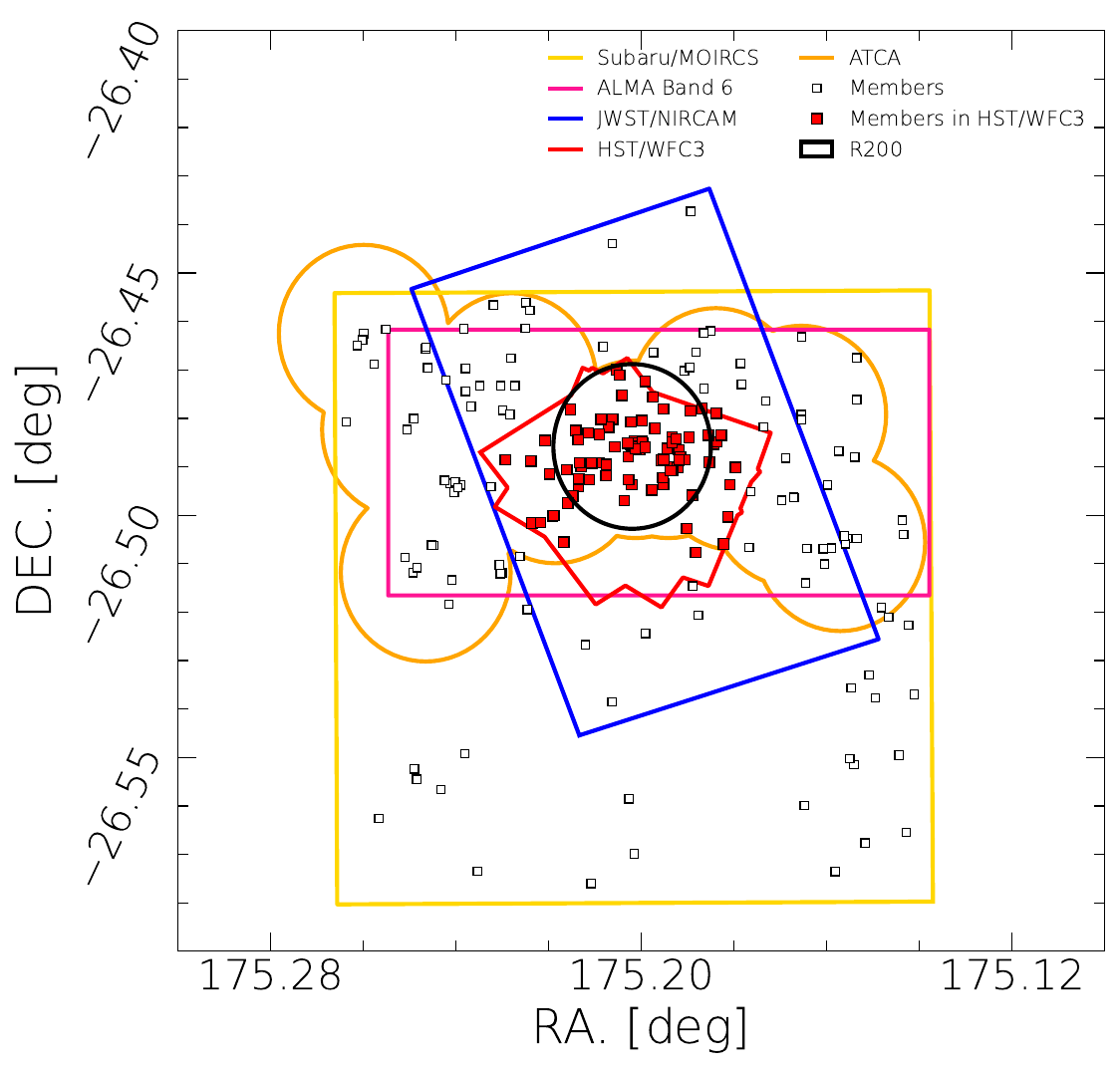}
    \caption{ {Footprints of the sky coverage of the main programs used to determine our member sample or protocluster galaxies within the spiderweb field. The telescopes and instrument utilized are HST/WFC3 (\citealt{Naufal2024ApJ...977...58N}), JWST/NIRCam (\citealt{Shimakawa2024ApJ...977...73S}; \citealt{Perez-Martinez_2024}), ALMA (\citealt{Zhang2024A&A...692A..22Z}), ATCA (\citealt{Jin2021A&A...652A..11J}), and Subaru/MOIRCS (\citealt{Koyama2013MNRAS.428.1551K}). The black circle represent the area enclosed by R200 as defined in \citet{Shimakawa2014MNRAS.441L...1S}. White and red squares respectively represent the full and restricted member sample, as described in Appendix~\ref{sec_app:footprint}.}
    \label{fig:footprint}}
\end{figure}

 {The member galaxy catalogue used in this work combines multiwavelength information from several follow-up programs targeting the Spiderweb protocluster. The main samples are based on data from HST/WFC3 (\citealt{Naufal2024ApJ...977...58N}), JWST/NIRCam (\citealt{Shimakawa2024ApJ...977...73S}; \citealt{Perez-Martinez_2024}), ALMA (\citealt{Zhang2024A&A...692A..22Z}), ATCA (\citealt{Jin2021A&A...652A..11J}), and Subaru/MOIRCS (\citealt{Koyama2013MNRAS.428.1551K}). These datasets do not all cover the same area, as shown in Fig.~\ref{fig:footprint}. Therefore, the spatial sampling of different galaxy populations is not perfectly uniform across the protocluster field. This uneven coverage could, in principle, affect the computation of local galaxy densities, particularly if specific populations are preferentially identified in regions with more complete follow-up. Such an effect could introduce unintended biases in the passive-density relation discussed in Sect.~\ref{sec:results_environ}.}

\begin{figure}[ht!]
\includegraphics[width=0.5\textwidth,clip]{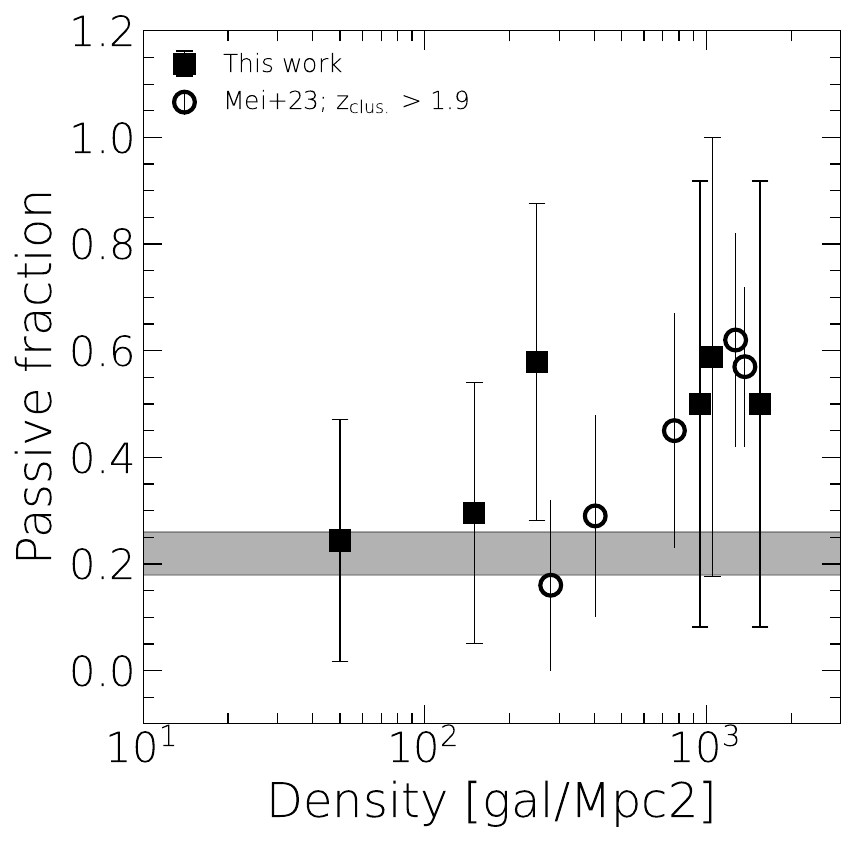}
    \caption{ {Passive-density relation after constraining our sample to objects lying within HST/WFC3 footprint as described in Appendix~\ref{sec_app:footprint}.}
    \label{fig:sigma3_hst_grism}}
\end{figure}

 {To assess the robustness of our results against this potential bias, we repeat the analysis after restricting the member sample to the innermost region of the protocluster, defined by the HST/WFC3 footprint. This region is also covered by the other main follow-up datasets considered here, providing a more homogeneous area over which to compute local density estimates. We then recompute the passive--density relation using this restricted sample, shown in Fig.~\ref{fig:sigma3_hst_grism}. We find that the resulting relation is largely consistent with that presented in Sect.~\ref{sec:results_environ}, maintaining a high passive fraction of $\sim50$-$60\%$ in the densest regions, while showing a slightly higher passive fraction at intermediate densities ($\sim 100$--$300~\mathrm{gal/Mpc^{-2}}$). This increase is likely a consequence of restricting the analysis to the innermost region of the protocluster, where the HST/WFC3 footprint overlaps with the ancillary datasets and where many of the passive members were originally identified (\citealt{Naufal2024ApJ...977...58N}).} 

\section{The environment and other properties}
\label{sec_app:environment}
\begin{figure}[ht!]
\includegraphics[width=0.5\textwidth,clip]{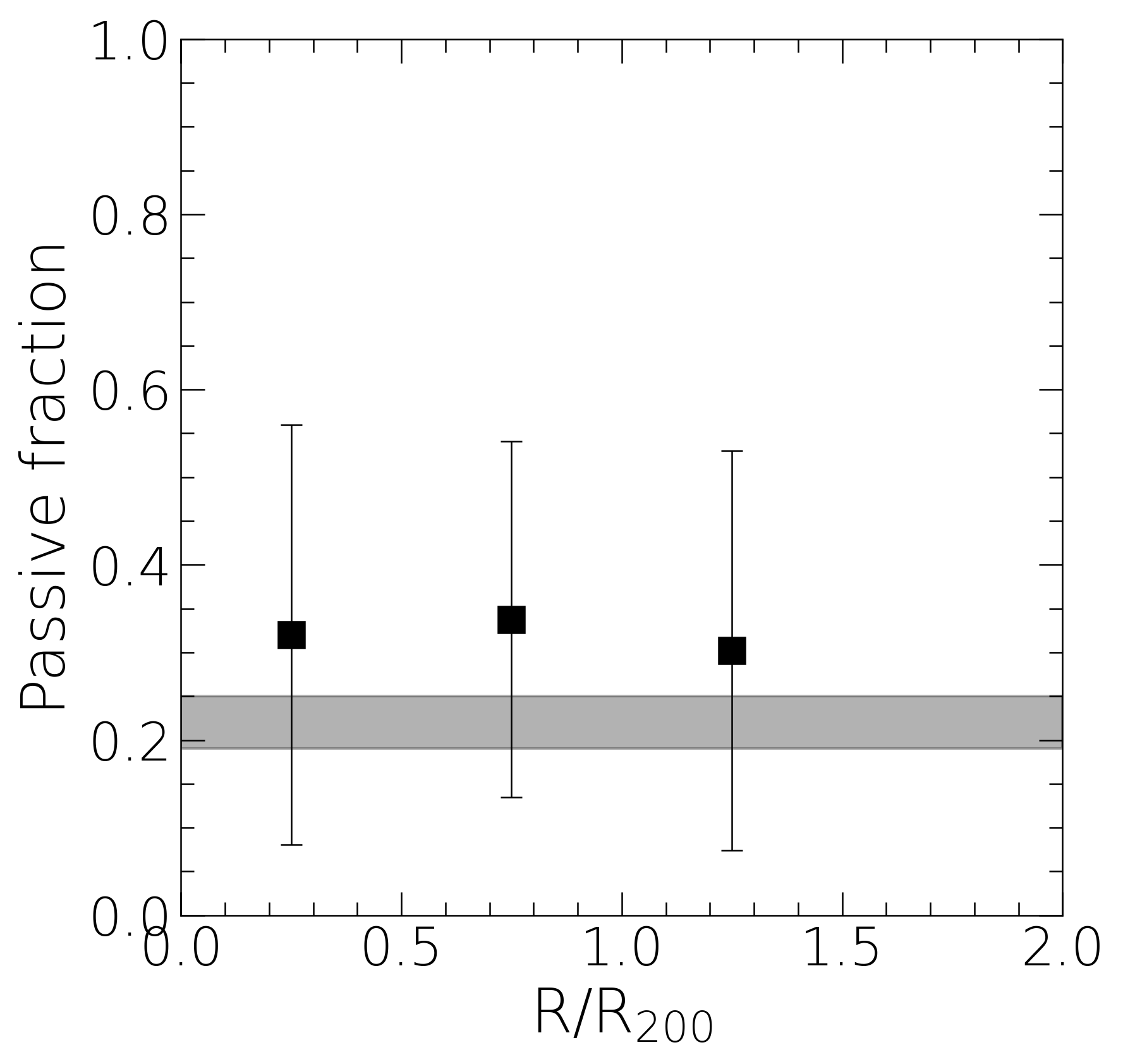}
    \caption{Fraction of passive galaxies as a function of distance from the radio galaxy PKS1138-262, normalised to R$_{200}$ as given in \citet{Shimakawa2024}. Grey horizontal shaded region shows field passive fraction \citep{yang2025_QGdensities}.
    \label{fig:RR200_passiveFraction}}
\end{figure}

\begin{figure*}[ht!]
\includegraphics[width=0.3\textwidth,clip]{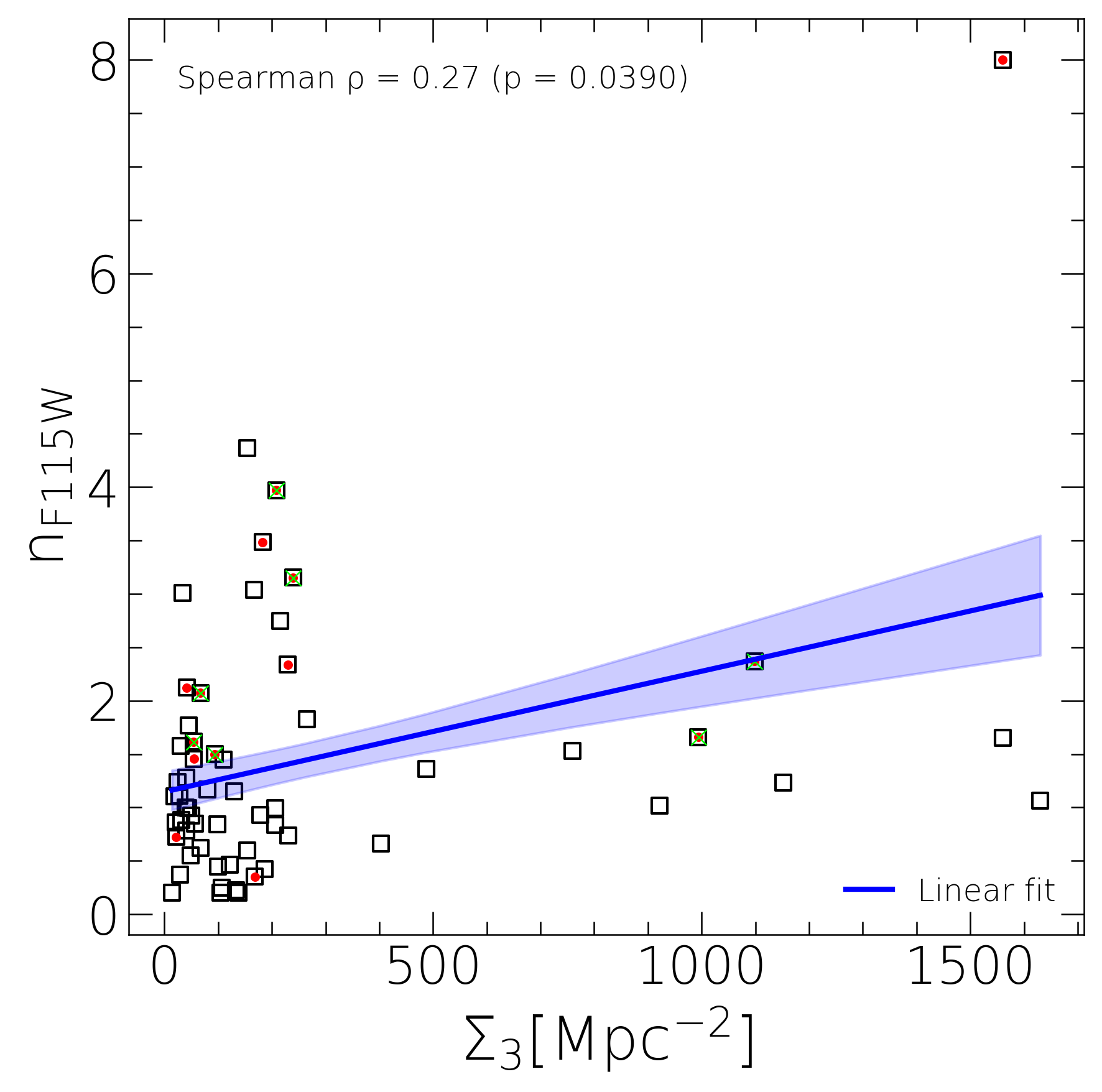}
\includegraphics[width=0.3\textwidth,clip]{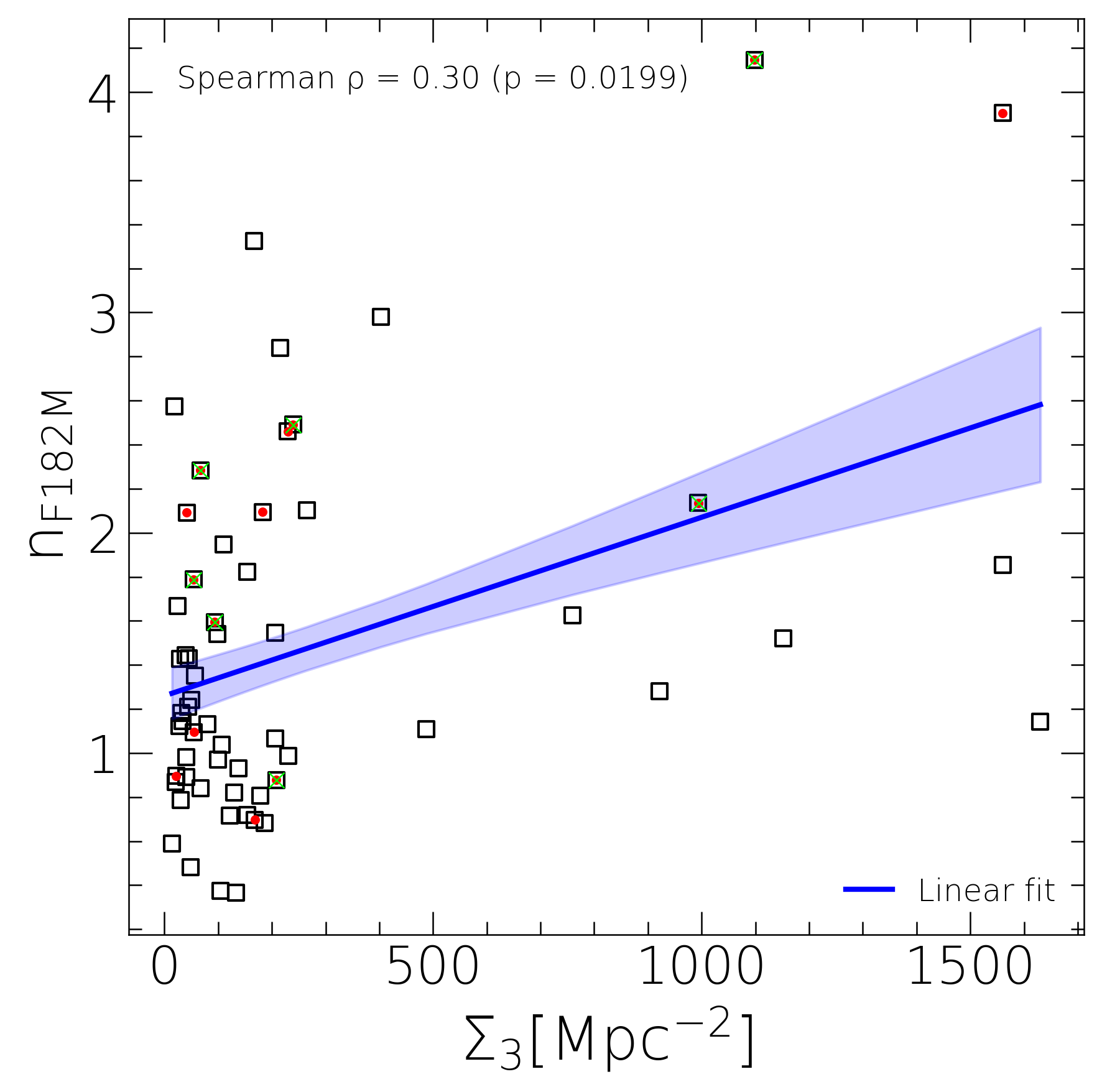}
\includegraphics[width=0.3\textwidth,clip]{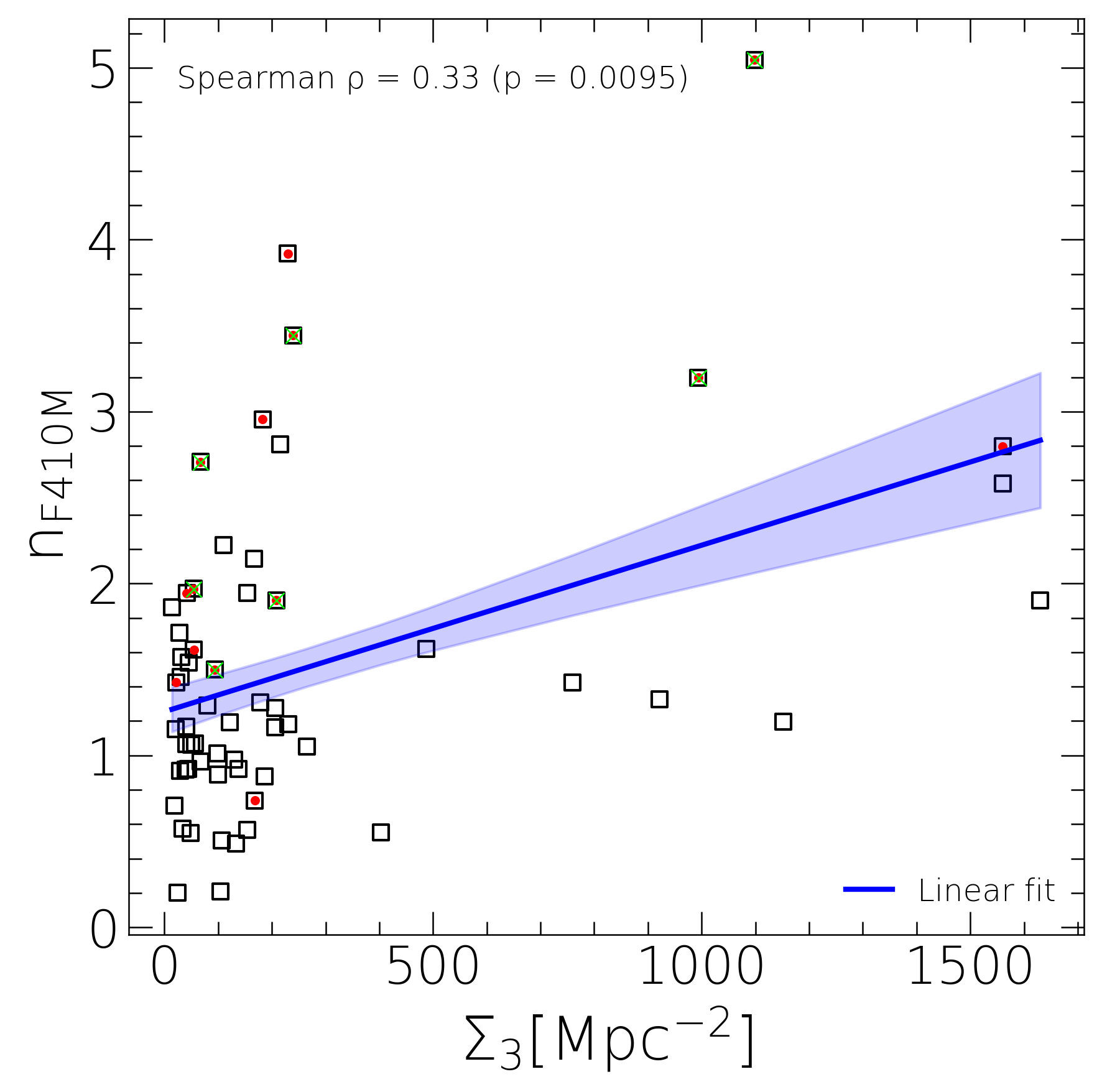}
\includegraphics[width=0.3\textwidth,clip]{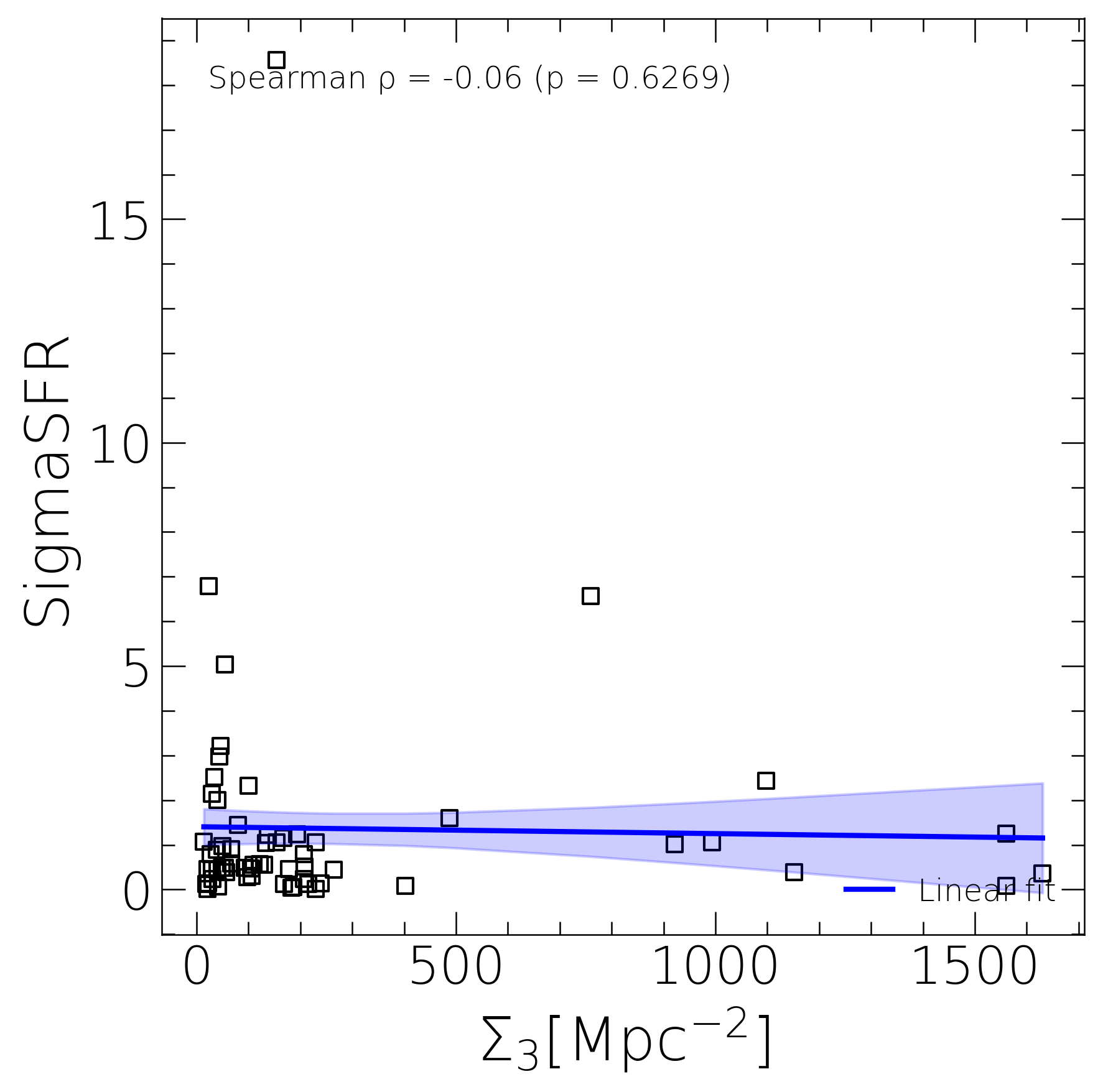}
\includegraphics[width=0.3\textwidth,clip]{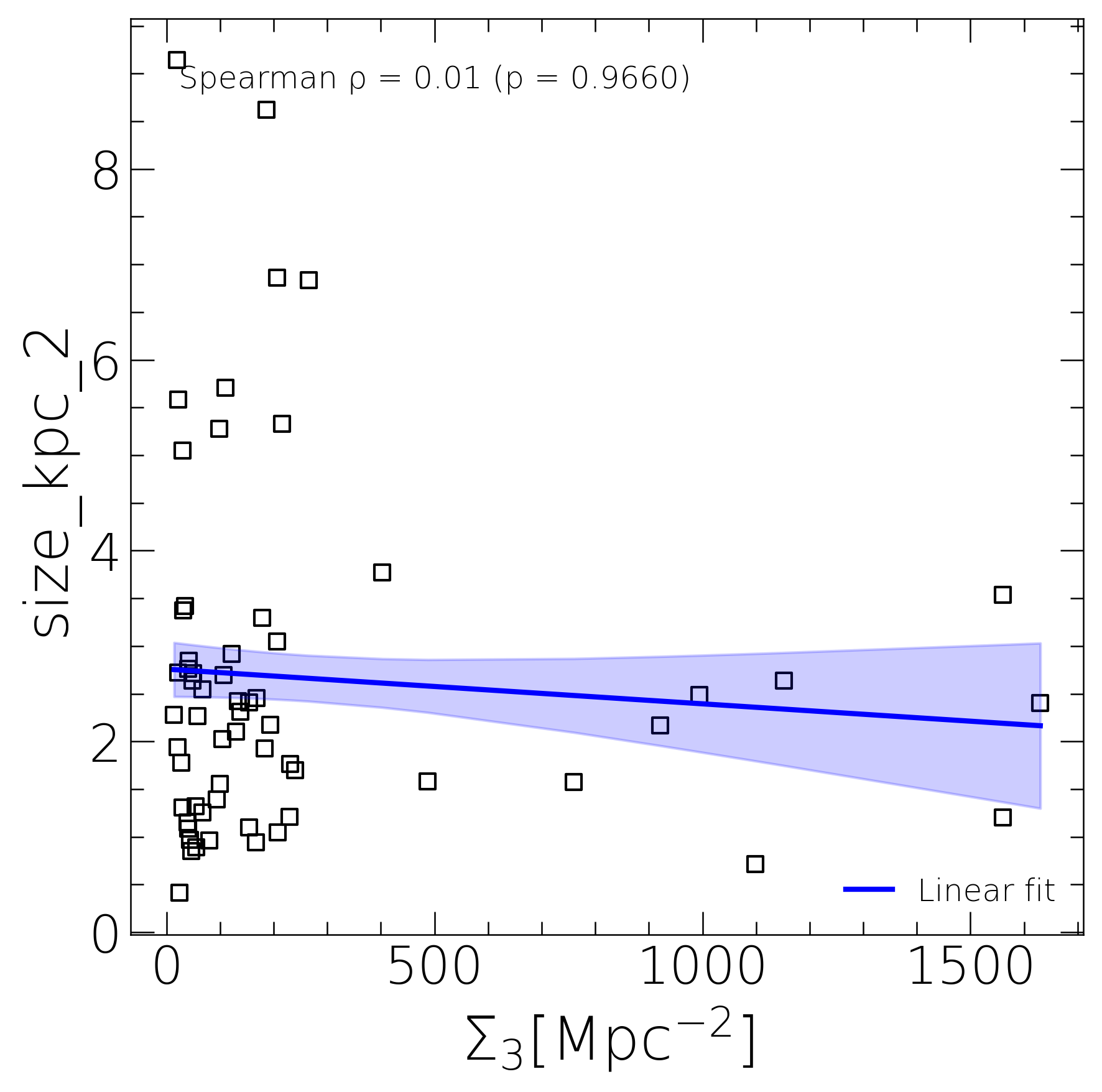}
    \caption{S\'ersic index as a function of galaxy density $\Sigma_3$. 
    \label{fig_app_sigma3_n}}
\end{figure*}

Figure \ref{fig:RR200_passiveFraction} shows the passive fraction as a function of the distance to Spiderweb galaxy PKS1138-262 normalised to $R_{200}$.
Figure \ref{fig_app_sigma3_n} shows the relation of local environment ( $\Sigma_3$) measurements and several parameters. 

\section{GALFIT-M results}
\begin{figure*}[ht!]
\centering
\fbox{\includegraphics[width=0.48\textwidth,clip]{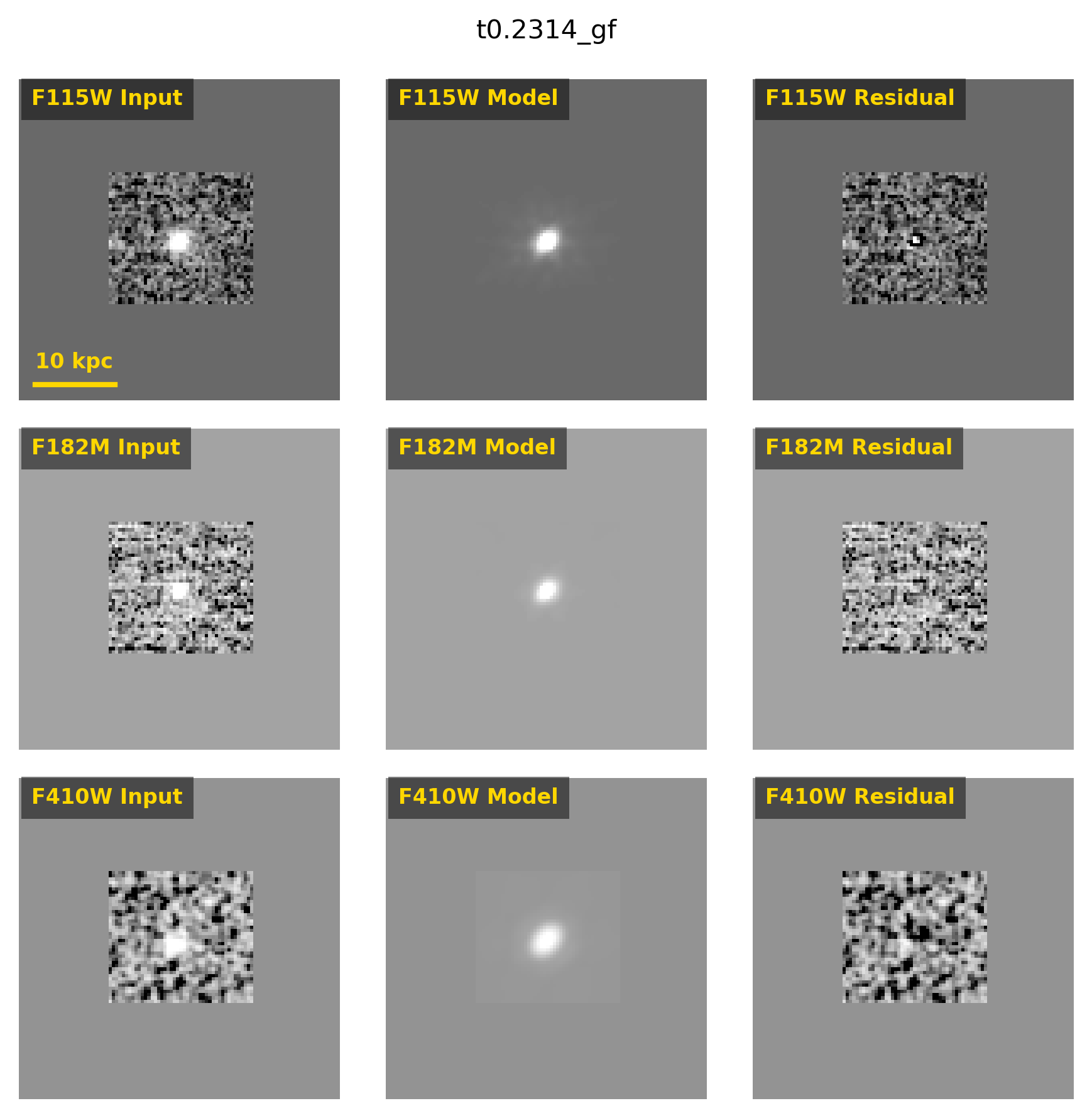}}
\fbox{\includegraphics[width=0.48\textwidth,clip]{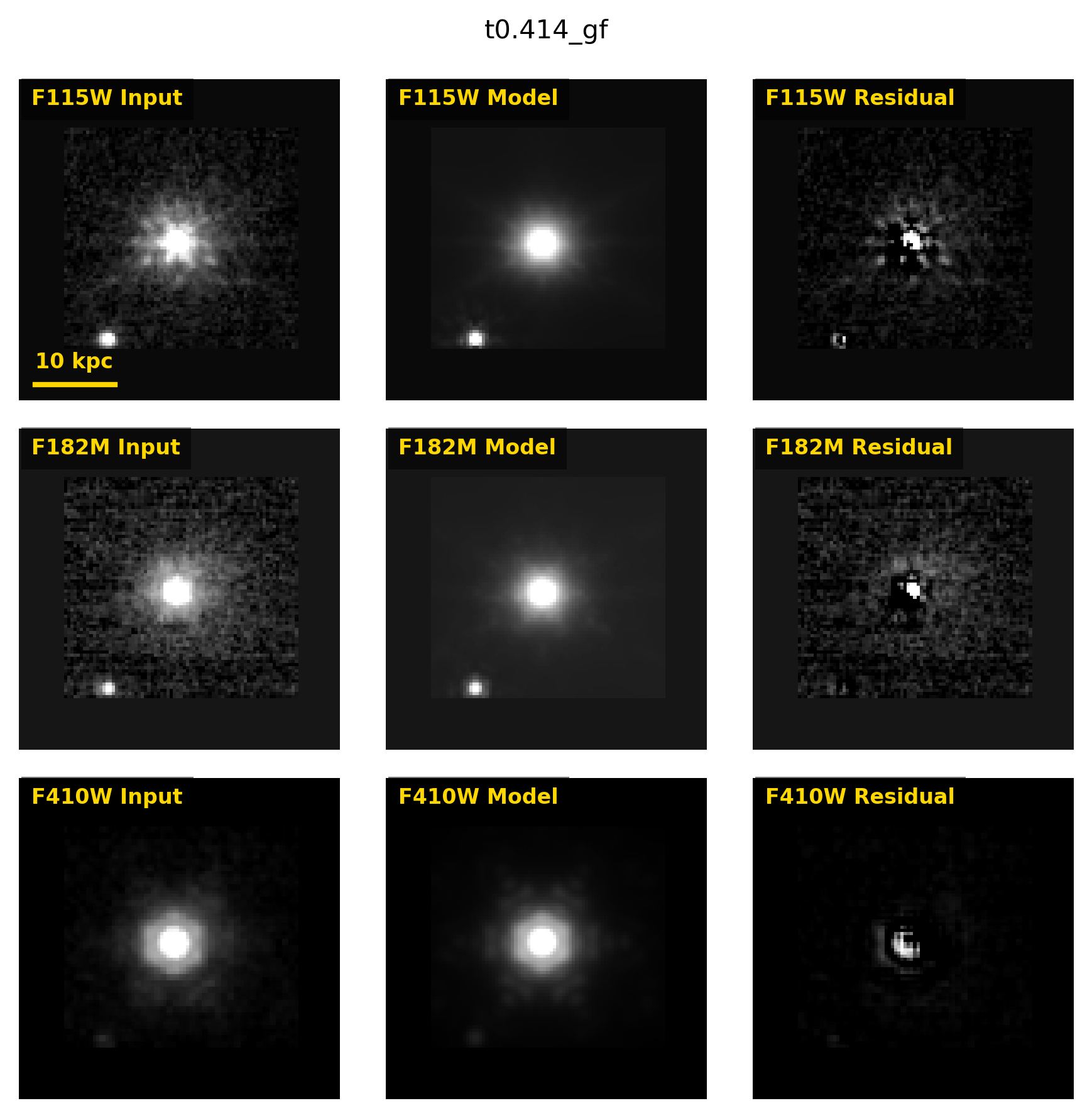}}
\caption{Results of the fitting using GALFIT-M for the two most compact sources.
\label{fig_appendix_compact}}
\end{figure*}

\begin{figure*}[ht!]
\centering
\fbox{\includegraphics[width=0.48\textwidth,clip]{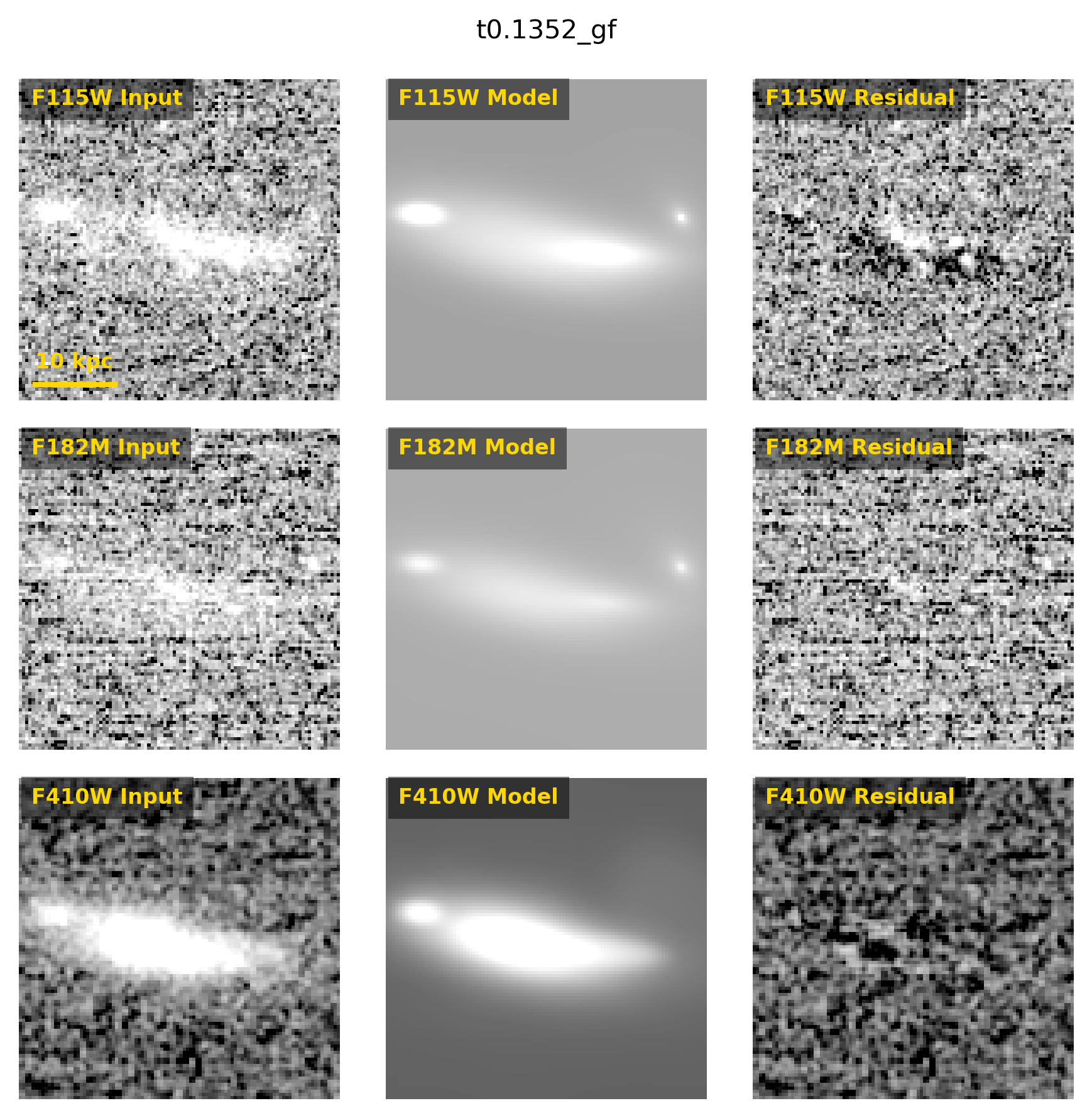}}
\fbox{\includegraphics[width=0.48\textwidth,clip]{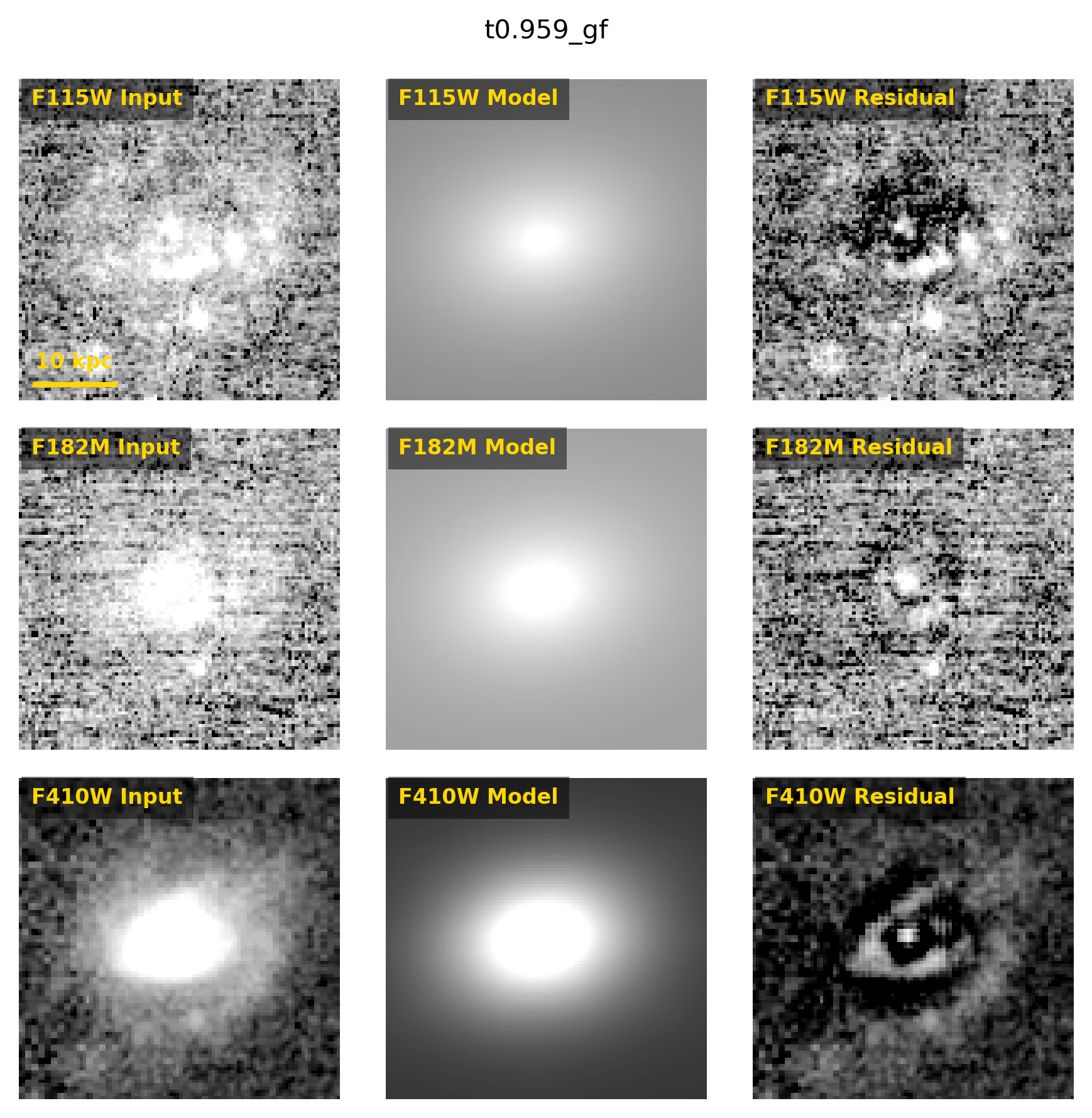}}
\caption{Results of the fitting using GALFIT-M for the two largest sources.
\label{fig_appendix_large}}
\end{figure*}

\begin{figure*}[ht!]
\centering
\fbox{\includegraphics[width=0.48\textwidth,clip]{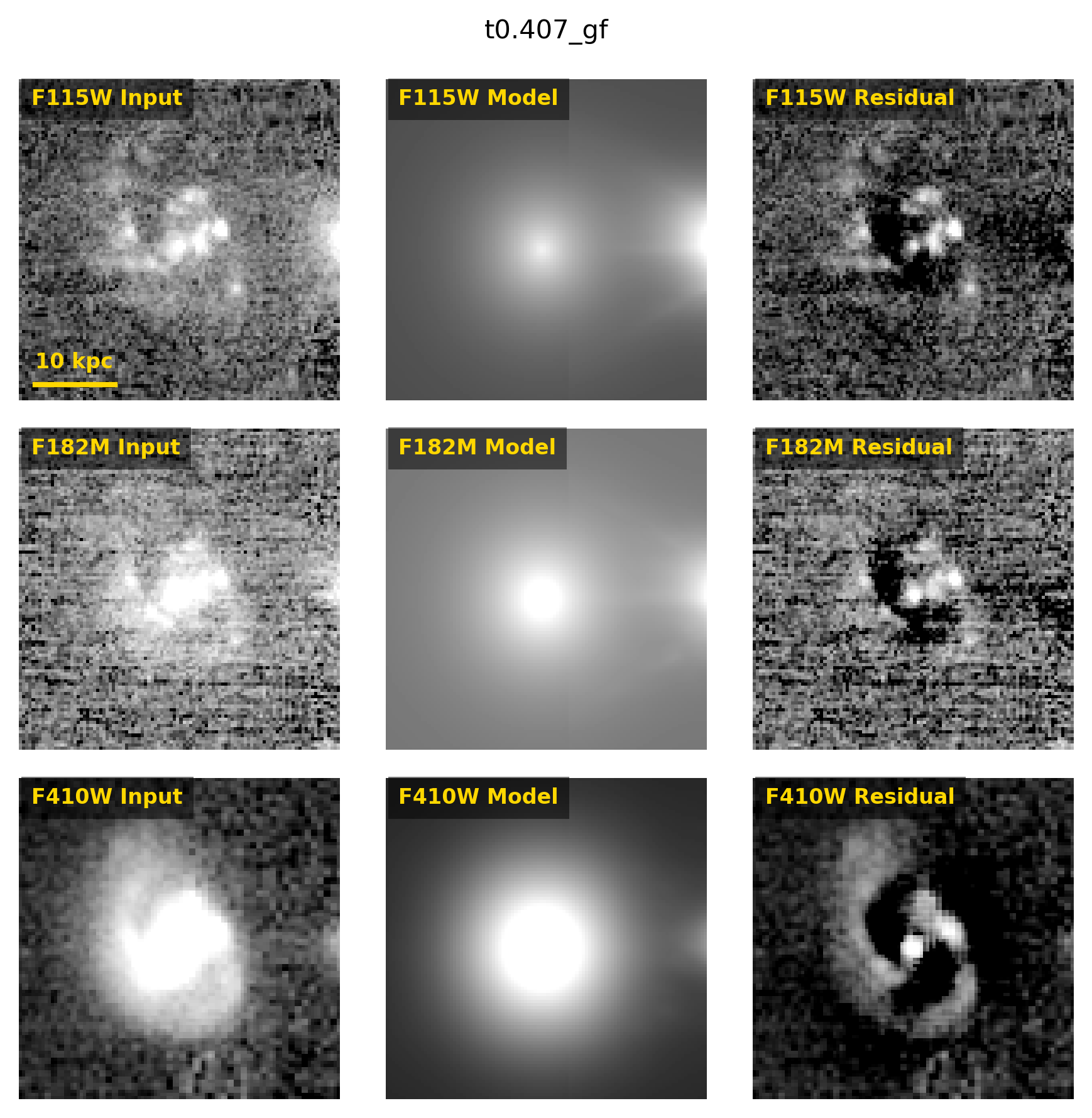}}
\fbox{\includegraphics[width=0.48\textwidth,clip]{t0.959_gf.png}}
\fbox{\includegraphics[width=0.48\textwidth,clip]{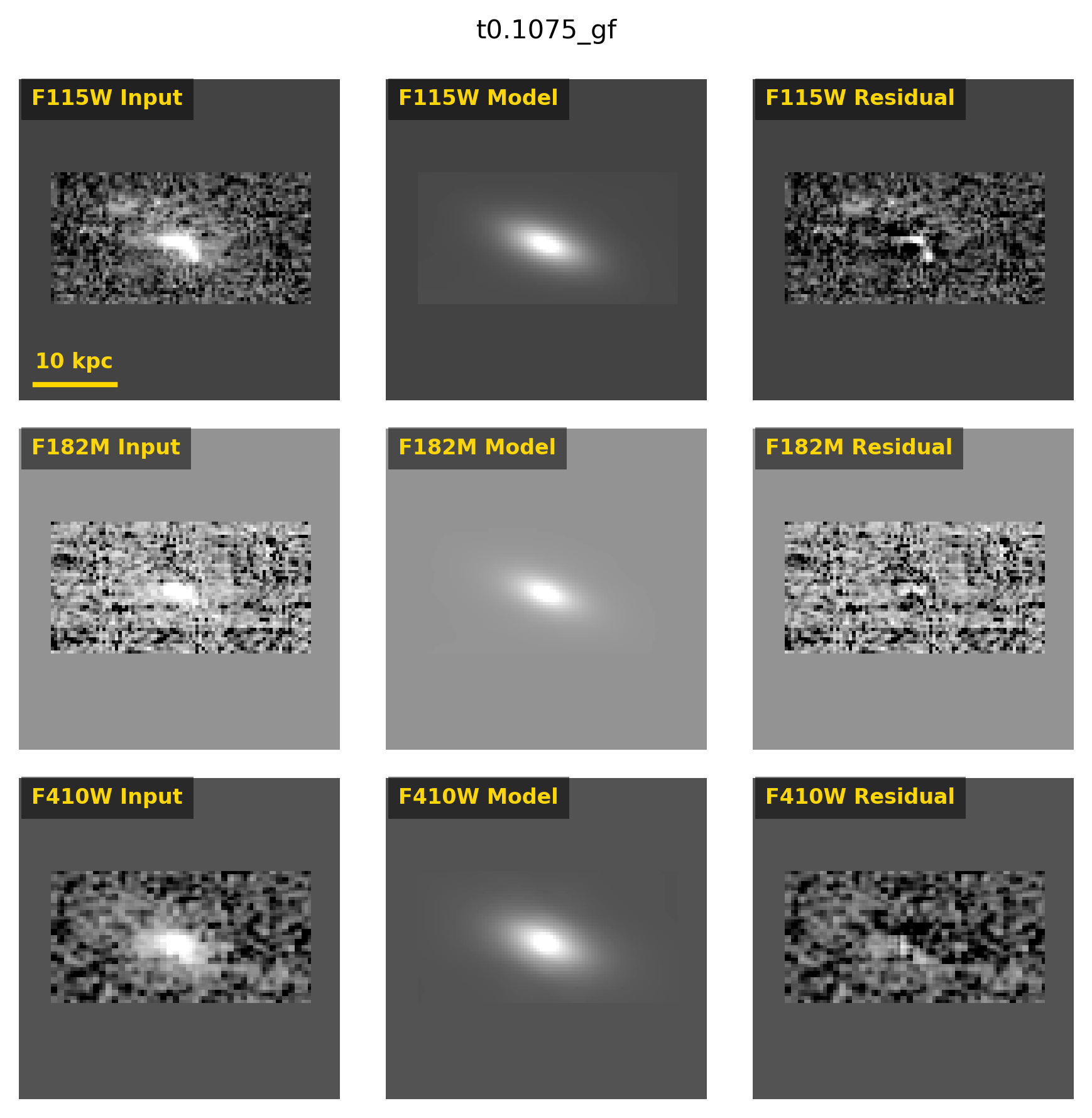}}
\fbox{\includegraphics[width=0.48\textwidth,clip]{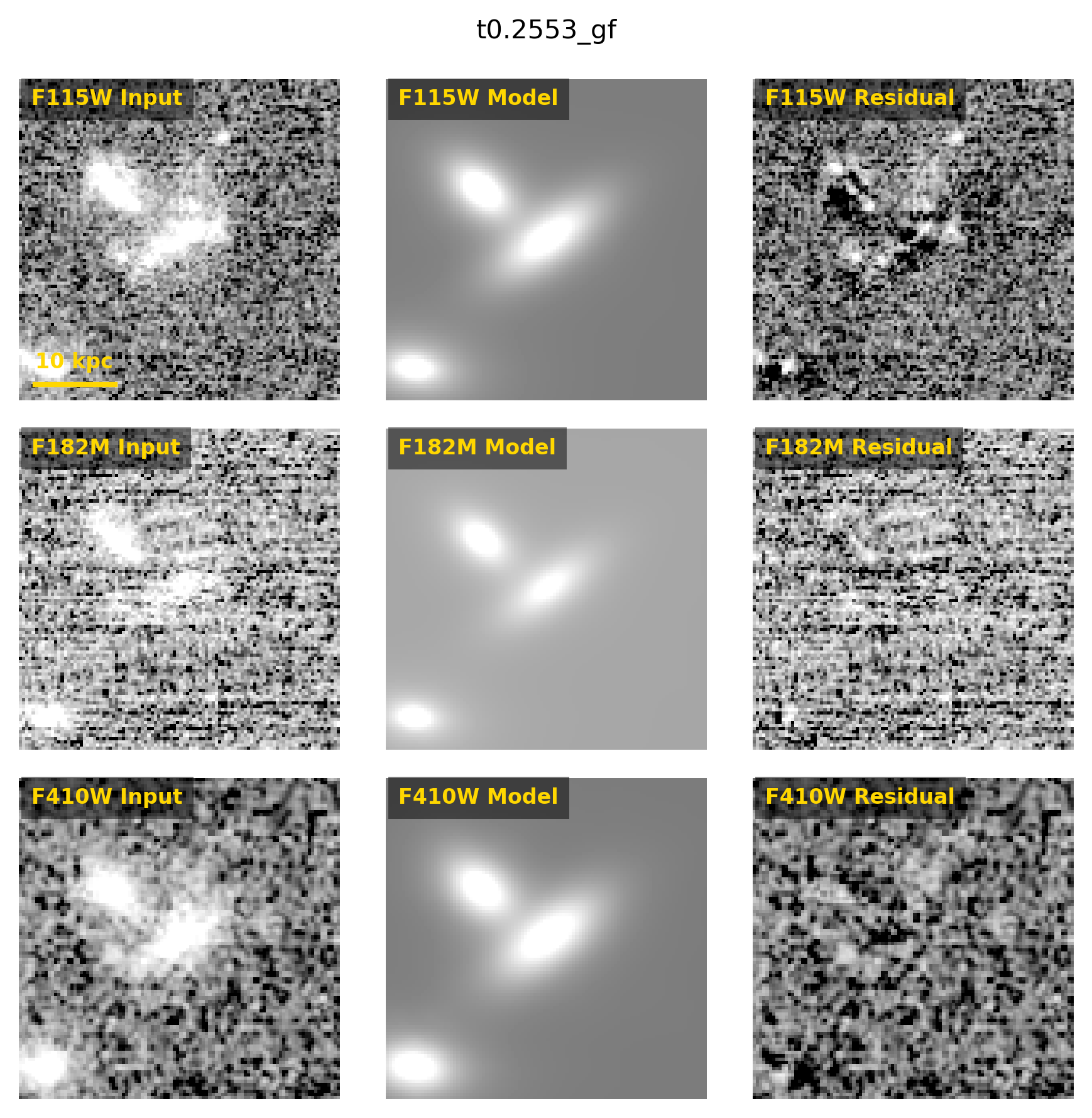}}
\caption{Results of the fitting using GALFIT-M for sources in common with \citet{Pérez-Martínez2023MNRAS.518.1707P}. These are: \idgala: 407  (or 902 in \citealt{Pérez-Martínez2023MNRAS.518.1707P}); \idgala: 959 (1300); \idgala: 1075 (1385); \idgala: 2553 (1071).
\label{fig_appendix_common}}
\end{figure*}

Here we show results from our fitting of selected galaxies. Figures \ref{fig_appendix_compact}--\ref{fig_appendix_common} are standarised to cover 100x100 pixles. The results from GALFIT-M varies from galaxy to galaxy. Thus, for some smaller objects the GALFIT-M cutout is smaller, and filled with "zeros". A 10 kpc scale bar is shown for each galaxy. 

\section{Summary Table}
Table \ref{tab:morph} resumes the results of our morphological analysis, including effective radii for each band, S\'ersic indices, and star formation surface density. As shown in simulations provided by \citet{Haussler2007ApJS..172..615H} the internal errors from GALFIT-M are underestimated. To the errors given in Table \ref{tab:morph} we have added the average error for S\'ersic index (0.46) and effective radii (0.16 px with the same pixel scale) from \citet{Haussler2007ApJS..172..615H}. This decision was made due to the fact that their errors does not show strong correlation with the surface brightness in the same range of our sample. 

\begin{center}

\onecolumn
\begin{longtable}{ccccccccc}
\caption{Morphological properties of sample galaxies studied in this work.}
\label{tab:morph}\\
\toprule
 ID &  GALA ID & $R_{\rm eff,F115W}$ & $R_{\rm eff,F182M}$ & $R_{\rm eff,F444W}$ & $n_{\rm F115W}$ & $n_{\rm F182M}$ & $n_{\rm F444W}$ & $\Sigma_{\rm SFR}$ \\
\midrule
\endfirsthead
\caption[]{Morphological properties of sample galaxies studied in this work.} \\
\toprule
 ID &  GALA ID & $R_{\rm eff,F115W}$ & $R_{\rm eff,F182M}$ & $R_{\rm eff,F444W}$ & $n_{\rm F115W}$ & $n_{\rm F182M}$ & $n_{\rm F444W}$ & $\Sigma_{\rm SFR}$ \\
\midrule
\endhead
\midrule
\multicolumn{9}{r}{{Continued on next page}} \\
\midrule
\endfoot

\bottomrule
\endlastfoot
 16 &       39 &     $2.71 \pm 0.02$ &     $2.59 \pm 0.02$ &     $2.20 \pm 0.01$ & $0.23 \pm 0.01$ & $0.29 \pm 0.01$ & $0.49 \pm 0.00$ &                0.9 \\
 17 &      119 &     $2.07 \pm 0.03$ &     $2.07 \pm 0.02$ &     $2.08 \pm 0.02$ & $0.94 \pm 0.02$ & $0.98 \pm 0.02$ & $1.13 \pm 0.02$ &                1.7 \\
 21 &     1352 &     $7.57 \pm 0.11$ &     $6.90 \pm 0.09$ &     $4.63 \pm 0.03$ & $0.43 \pm 0.02$ & $0.53 \pm 0.01$ & $0.89 \pm 0.01$ &                0.1 \\
 22 &     1388 &     $3.53 \pm 0.18$ &     $3.39 \pm 0.14$ &     $2.93 \pm 0.05$ & $1.35 \pm 0.07$ & $1.21 \pm 0.06$ & $0.72 \pm 0.02$ &                0.9 \\
 23 &     1381 &     $2.39 \pm 0.03$ &     $2.57 \pm 0.03$ &     $3.16 \pm 0.09$ & $0.20 \pm 0.01$ & $0.41 \pm 0.01$ & $1.12 \pm 0.05$ &                0.8 \\
 24 &      272 &     $3.55 \pm 0.08$ &     $3.49 \pm 0.07$ &     $3.27 \pm 0.05$ & $1.25 \pm 0.03$ & $1.22 \pm 0.03$ & $1.09 \pm 0.03$ &                0.3 \\
 26 &      273 &     $2.65 \pm 0.04$ &     $2.71 \pm 0.03$ &     $2.93 \pm 0.04$ & $0.55 \pm 0.02$ & $0.55 \pm 0.01$ & $0.55 \pm 0.02$ &                0.5 \\
 27 &      308 &     $1.93 \pm 0.04$ &     $1.91 \pm 0.03$ &     $1.85 \pm 0.01$ & $2.85 \pm 0.05$ & $2.98 \pm 0.04$ & $3.42 \pm 0.02$ &                0.1 \\
 28 &      337 &     $1.34 \pm 0.01$ &     $1.28 \pm 0.01$ &     $1.05 \pm 0.00$ & $1.61 \pm 0.01$ & $1.70 \pm 0.01$ & $1.97 \pm 0.00$ &                0.5 \\
 37 &     2007 &     $1.48 \pm 0.04$ &     $1.50 \pm 0.04$ &     $1.59 \pm 0.05$ & $0.89 \pm 0.04$ & $0.93 \pm 0.04$ & $1.08 \pm 0.07$ &                0.9 \\
 38 &      458 &     $1.21 \pm 0.04$ &     $1.32 \pm 0.03$ &     $1.69 \pm 0.01$ & $1.73 \pm 0.06$ & $1.77 \pm 0.05$ & $1.91 \pm 0.01$ &                0.3 \\
 39 &      449 &     $4.47 \pm 0.06$ &     $4.09 \pm 0.04$ &     $2.78 \pm 0.01$ & $1.50 \pm 0.02$ & $1.73 \pm 0.01$ & $2.53 \pm 0.01$ &                0.9 \\
 41 &      407 &     $5.52 \pm 0.02$ &     $5.17 \pm 0.02$ &     $3.96 \pm 0.00$ & $1.23 \pm 0.01$ & $1.19 \pm 0.00$ & $1.06 \pm 0.00$ &                0.8 \\
 42 &      442 &     $1.53 \pm 0.01$ &     $1.53 \pm 0.01$ &     $1.51 \pm 0.01$ & $1.53 \pm 0.01$ & $1.51 \pm 0.01$ & $1.43 \pm 0.01$ &                7.0 \\
 43 &      414 &     $0.57 \pm 0.01$ &     $0.58 \pm 0.00$ &     $0.64 \pm 0.01$ & $2.50 \pm 0.04$ & $3.10 \pm 0.04$ & $5.13 \pm 0.10$ &                3.7 \\
 44 &      319 &     $1.00 \pm 0.02$ &     $0.95 \pm 0.01$ &     $0.79 \pm 0.00$ & $1.24 \pm 0.03$ & $1.33 \pm 0.03$ & $1.61 \pm 0.02$ &                4.4 \\
 45 &      410 &     $1.95 \pm 0.00$ &     $1.89 \pm 0.00$ &     $1.70 \pm 0.00$ & $1.07 \pm 0.00$ & $1.13 \pm 0.00$ & $1.32 \pm 0.00$ &                1.3 \\
 46 &      301 &     $2.89 \pm 0.16$ &     $2.58 \pm 0.13$ &     $1.51 \pm 0.01$ & $1.53 \pm 0.09$ & $1.53 \pm 0.07$ & $1.50 \pm 0.01$ &                0.1 \\
 47 &     1774 &     $2.68 \pm 0.10$ &     $2.54 \pm 0.08$ &     $2.05 \pm 0.04$ & $1.11 \pm 0.05$ & $1.08 \pm 0.04$ & $0.97 \pm 0.04$ &                0.4 \\
 57 &     2553 &     $3.45 \pm 0.01$ &     $3.43 \pm 0.01$ &     $3.37 \pm 0.01$ & $0.89 \pm 0.01$ & $0.91 \pm 0.00$ & $0.99 \pm 0.01$ &                0.7 \\
 58 &      778 &     $2.59 \pm 0.01$ &     $2.50 \pm 0.00$ &     $2.20 \pm 0.00$ & $0.64 \pm 0.00$ & $0.71 \pm 0.00$ & $0.97 \pm 0.00$ &                0.9 \\
 61 &     2927 &     $2.22 \pm 0.08$ &     $2.14 \pm 0.06$ &     $1.86 \pm 0.03$ & $0.86 \pm 0.05$ & $0.93 \pm 0.04$ & $1.18 \pm 0.03$ &                0.4 \\
 62 &      811 &     $3.43 \pm 0.14$ &     $3.11 \pm 0.11$ &     $2.03 \pm 0.01$ & $2.12 \pm 0.08$ & $2.08 \pm 0.06$ & $1.95 \pm 0.02$ &                0.1 \\
 63 &      866 &     $4.21 \pm 0.18$ &     $3.96 \pm 0.14$ &     $3.11 \pm 0.01$ & $2.70 \pm 0.08$ & $2.22 \pm 0.06$ & $0.58 \pm 0.01$ &                1.9 \\
 64 &     2904 &     $7.14 \pm 0.16$ &     $6.37 \pm 0.12$ &     $3.77 \pm 0.03$ & $0.32 \pm 0.03$ & $0.61 \pm 0.02$ & $1.60 \pm 0.02$ &                0.3 \\
 65 &      699 &     $0.96 \pm 0.02$ &     $0.94 \pm 0.02$ &     $0.88 \pm 0.04$ & $1.79 \pm 0.08$ & $1.73 \pm 0.08$ & $1.55 \pm 0.19$ &                2.7 \\
 68 &      839 &     $2.38 \pm 0.02$ &     $2.32 \pm 0.02$ &     $2.11 \pm 0.01$ & $0.62 \pm 0.02$ & $0.60 \pm 0.01$ & $0.56 \pm 0.01$ &                1.1 \\
 69 &      959 &     $8.45 \pm 0.16$ &     $7.46 \pm 0.12$ &     $4.07 \pm 0.01$ & $1.08 \pm 0.02$ & $1.12 \pm 0.02$ & $1.28 \pm 0.00$ &                0.2 \\
 70 &      948 &     $3.20 \pm 0.02$ &     $3.04 \pm 0.01$ &     $2.50 \pm 0.01$ & $1.07 \pm 0.01$ & $1.09 \pm 0.01$ & $1.16 \pm 0.00$ &                0.8 \\
 71 &     1047 &     $3.63 \pm 0.02$ &     $3.44 \pm 0.02$ &     $2.81 \pm 0.01$ & $0.87 \pm 0.01$ & $1.03 \pm 0.01$ & $1.58 \pm 0.01$ &                0.2 \\
 73 &     1075 &     $2.06 \pm 0.05$ &     $2.09 \pm 0.05$ &     $2.20 \pm 0.06$ & $1.30 \pm 0.04$ & $1.38 \pm 0.03$ & $1.63 \pm 0.07$ &                0.9 \\
 74 &     3030 &     $2.64 \pm 0.02$ &     $2.74 \pm 0.02$ &     $3.08 \pm 0.03$ & $0.26 \pm 0.01$ & $0.33 \pm 0.01$ & $0.60 \pm 0.02$ &                0.3 \\
 75 &      628 &     $3.86 \pm 0.03$ &     $3.56 \pm 0.03$ &     $2.54 \pm 0.01$ & $0.95 \pm 0.01$ & $1.16 \pm 0.01$ & $1.86 \pm 0.01$ &                    \\
 76 &      651 &     $0.19 \pm 0.00$ &     $0.22 \pm 0.00$ &     $0.33 \pm 0.02$ & $0.70 \pm 0.05$ & $0.88 \pm 0.11$ & $1.47 \pm 0.48$ &                    \\
 77 &      599 &     $2.38 \pm 0.04$ &     $2.65 \pm 0.04$ &     $3.56 \pm 0.09$ & $1.09 \pm 0.02$ & $1.34 \pm 0.02$ & $2.18 \pm 0.05$ &                0.3 \\
 80 &      880 &     $4.68 \pm 0.10$ &     $4.10 \pm 0.08$ &     $2.12 \pm 0.01$ & $0.86 \pm 0.03$ & $0.96 \pm 0.02$ & $1.31 \pm 0.01$ &                0.3 \\
 82 &     2383 &     $3.50 \pm 0.00$ &     $3.28 \pm 0.00$ &     $2.51 \pm 0.00$ & $0.47 \pm 0.00$ & $0.64 \pm 0.00$ & $1.20 \pm 0.00$ &                0.5 \\
 83 &      626 &     $2.56 \pm 0.05$ &     $2.44 \pm 0.04$ &     $2.05 \pm 0.02$ & $1.71 \pm 0.03$ & $2.05 \pm 0.02$ & $3.22 \pm 0.02$ &                1.1 \\
 86 &      743 &     $2.25 \pm 0.38$ &     $2.20 \pm 0.29$ &     $2.04 \pm 0.04$ & $2.72 \pm 0.36$ & $2.58 \pm 0.28$ & $2.11 \pm 0.07$ &                4.6 \\
 88 &      586 &     $1.74 \pm 0.03$ &     $1.64 \pm 0.03$ &     $1.28 \pm 0.00$ & $0.72 \pm 0.04$ & $0.76 \pm 0.03$ & $0.89 \pm 0.01$ &                2.1 \\
 89 &     2366 &     $0.67 \pm 0.03$ &     $0.80 \pm 0.03$ &     $1.25 \pm 0.07$ & $3.03 \pm 0.16$ & $2.84 \pm 0.13$ & $2.17 \pm 0.21$ &                1.6 \\
 90 &     1016 &     $2.27 \pm 0.05$ &     $2.36 \pm 0.04$ &     $2.68 \pm 0.01$ & $0.43 \pm 0.03$ & $0.49 \pm 0.03$ & $0.69 \pm 0.01$ &                0.1 \\
 91 &     1041 &     $2.06 \pm 0.04$ &     $1.95 \pm 0.03$ &     $1.57 \pm 0.03$ & $0.75 \pm 0.03$ & $0.85 \pm 0.02$ & $1.19 \pm 0.03$ &                0.9 \\
 93 &     2781 &     $1.58 \pm 0.05$ &     $1.56 \pm 0.04$ &     $1.49 \pm 0.03$ & $0.85 \pm 0.04$ & $0.87 \pm 0.03$ & $0.93 \pm 0.04$ &                1.1 \\
 94 &      501 &     $2.65 \pm 0.00$ &     $2.60 \pm 0.00$ &     $2.43 \pm 0.00$ & $0.20 \pm 0.00$ & $0.37 \pm 0.00$ & $0.93 \pm 0.00$ &                1.0 \\
 96 &     1764 &     $0.93 \pm 0.03$ &     $0.92 \pm 0.02$ &     $0.88 \pm 0.03$ & $0.70 \pm 0.05$ & $0.67 \pm 0.04$ & $0.56 \pm 0.08$ &                1.3 \\
 97 &     2961 &     $2.51 \pm 0.21$ &     $2.21 \pm 0.16$ &     $1.20 \pm 0.02$ & $1.05 \pm 0.11$ & $1.20 \pm 0.09$ & $1.75 \pm 0.04$ &                0.5 \\
 98 &     2292 &     $3.73 \pm 0.17$ &     $3.37 \pm 0.13$ &     $2.15 \pm 0.02$ & $1.11 \pm 0.07$ & $1.10 \pm 0.05$ & $1.10 \pm 0.02$ &                0.2 \\
 99 &     1937 &     $2.18 \pm 0.03$ &     $2.16 \pm 0.03$ &     $2.11 \pm 0.04$ & $0.20 \pm 0.02$ & $0.23 \pm 0.02$ & $0.32 \pm 0.04$ &                0.4 \\
100 &     2112 &     $0.89 \pm 0.02$ &     $0.88 \pm 0.01$ &     $0.85 \pm 0.01$ & $1.13 \pm 0.04$ & $1.11 \pm 0.03$ & $1.06 \pm 0.06$ &                1.7 \\
102 &     1149 &     $7.69 \pm 0.96$ &     $6.73 \pm 0.75$ &     $3.46 \pm 0.11$ & $2.62 \pm 0.19$ & $2.67 \pm 0.14$ & $2.84 \pm 0.08$ &                0.1 \\
103 &      664 &     $1.98 \pm 0.03$ &     $1.95 \pm 0.02$ &     $1.86 \pm 0.02$ & $1.29 \pm 0.02$ & $1.30 \pm 0.02$ & $1.35 \pm 0.02$ &                1.0 \\
105 &      282 &     $1.01 \pm 0.03$ &     $1.03 \pm 0.02$ &     $1.10 \pm 0.02$ & $0.78 \pm 0.04$ & $0.87 \pm 0.03$ & $1.17 \pm 0.04$ &                2.2 \\
106 &      823 &     $1.06 \pm 0.04$ &     $1.03 \pm 0.03$ &     $0.90 \pm 0.01$ & $2.14 \pm 0.08$ & $2.26 \pm 0.07$ & $2.69 \pm 0.04$ &                0.9 \\
108 &     2314 &     $0.36 \pm 0.01$ &     $0.47 \pm 0.01$ &     $0.85 \pm 0.04$ & $1.15 \pm 0.09$ & $0.93 \pm 0.08$ & $0.20 \pm 0.22$ &                5.2 \\
110 &      373 &     $3.35 \pm 0.03$ &     $3.04 \pm 0.03$ &     $1.98 \pm 0.01$ & $0.73 \pm 0.01$ & $0.89 \pm 0.01$ & $1.43 \pm 0.01$ &                0.0 \\
112 &     1530 &     $1.24 \pm 0.01$ &     $1.24 \pm 0.01$ &     $1.22 \pm 0.01$ & $0.99 \pm 0.01$ & $0.98 \pm 0.01$ & $0.94 \pm 0.02$ &                0.8 \\
119 &      435 &     $6.94 \pm 0.01$ &     $6.68 \pm 0.00$ &     $5.82 \pm 0.01$ & $1.19 \pm 0.00$ & $1.22 \pm 0.00$ & $1.34 \pm 0.00$ &                    \\
120 &      705 &     $1.00 \pm 0.01$ &     $1.00 \pm 0.01$ &     $0.97 \pm 0.00$ & $1.60 \pm 0.02$ & $1.56 \pm 0.02$ & $1.42 \pm 0.01$ &                    \\
121 &       64 &     $6.13 \pm 0.16$ &     $5.56 \pm 0.13$ &     $3.63 \pm 0.06$ & $1.53 \pm 0.04$ & $1.70 \pm 0.03$ & $2.28 \pm 0.03$ &                0.6 \\
124 &     1325 &     $3.01 \pm 0.05$ &     $2.93 \pm 0.04$ &     $2.66 \pm 0.01$ & $0.45 \pm 0.02$ & $0.44 \pm 0.01$ & $0.40 \pm 0.00$ &                    \\
126 &       90 &     $1.82 \pm 0.06$ &     $1.79 \pm 0.05$ &     $1.70 \pm 0.01$ & $0.91 \pm 0.06$ & $0.93 \pm 0.04$ & $1.03 \pm 0.02$ &                    \\
127 &      166 &     $4.24 \pm 0.08$ &     $3.98 \pm 0.06$ &     $3.10 \pm 0.03$ & $2.04 \pm 0.03$ & $1.97 \pm 0.02$ & $1.71 \pm 0.02$ &                    \\
128 &     1657 &     $4.05 \pm 0.00$ &     $3.93 \pm 0.00$ &     $3.51 \pm 0.00$ & $0.50 \pm 0.00$ & $0.43 \pm 0.00$ & $0.22 \pm 0.00$ &                    \\
129 &      324 &     $0.85 \pm 0.01$ &     $0.88 \pm 0.01$ &     $0.99 \pm 0.01$ & $1.07 \pm 0.03$ & $1.23 \pm 0.03$ & $1.78 \pm 0.07$ &                    \\
130 &     1854 &     $0.91 \pm 0.04$ &     $0.93 \pm 0.03$ &     $1.01 \pm 0.03$ & $1.06 \pm 0.08$ & $1.09 \pm 0.06$ & $1.19 \pm 0.09$ &                    \\
131 &     1958 &     $2.78 \pm 0.32$ &     $2.60 \pm 0.25$ &     $1.98 \pm 0.06$ & $1.65 \pm 0.19$ & $1.90 \pm 0.15$ & $2.76 \pm 0.07$ &                    \\
133 &      436 &     $2.99 \pm 0.02$ &     $2.87 \pm 0.02$ &     $2.45 \pm 0.01$ & $0.78 \pm 0.01$ & $0.74 \pm 0.01$ & $0.61 \pm 0.01$ &                    \\
134 &      425 &     $1.53 \pm 0.00$ &     $1.57 \pm 0.00$ &     $1.71 \pm 0.01$ & $0.96 \pm 0.00$ & $1.00 \pm 0.00$ & $1.11 \pm 0.01$ &                    \\
135 &     2035 &     $0.30 \pm 0.02$ &     $0.47 \pm 0.06$ &     $1.04 \pm 0.27$ & $2.89 \pm 0.22$ & $2.92 \pm 0.25$ & $3.02 \pm 0.72$ &                    \\
138 &     2313 &     $2.51 \pm 0.15$ &     $2.31 \pm 0.11$ &     $1.61 \pm 0.01$ & $0.33 \pm 0.10$ & $0.48 \pm 0.08$ & $0.97 \pm 0.02$ &                    \\
139 &     2318 &     $0.80 \pm 0.07$ &     $0.75 \pm 0.06$ &     $0.58 \pm 0.03$ & $0.97 \pm 0.18$ & $1.04 \pm 0.14$ & $1.29 \pm 0.23$ &                    \\
140 &     2291 &     $0.74 \pm 0.00$ &     $0.80 \pm 0.00$ &     $0.98 \pm 0.00$ & $0.87 \pm 0.01$ & $0.72 \pm 0.01$ & $0.20 \pm 0.01$ &                    \\
141 &     1148 &     $1.75 \pm 0.06$ &     $1.63 \pm 0.05$ &     $1.23 \pm 0.02$ & $1.43 \pm 0.06$ & $1.43 \pm 0.04$ & $1.42 \pm 0.03$ &                    \\
142 &     3295 &     $3.96 \pm 0.24$ &     $3.52 \pm 0.18$ &     $1.99 \pm 0.04$ & $0.21 \pm 0.09$ & $0.24 \pm 0.07$ & $0.35 \pm 0.03$ &                    \\
143 &     1007 &     $1.47 \pm 0.02$ &     $1.43 \pm 0.02$ &     $1.31 \pm 0.01$ & $0.84 \pm 0.02$ & $0.87 \pm 0.02$ & $0.96 \pm 0.01$ &                    \\
145 &     1017 &     $2.20 \pm 0.02$ &     $2.19 \pm 0.01$ &     $2.15 \pm 0.02$ & $0.34 \pm 0.01$ & $0.37 \pm 0.01$ & $0.48 \pm 0.01$ &                    \\
146 &     2517 &     $3.22 \pm 0.33$ &     $2.89 \pm 0.25$ &     $1.78 \pm 0.04$ & $0.73 \pm 0.13$ & $0.77 \pm 0.10$ & $0.93 \pm 0.04$ &                    \\
154 &     2588 &     $1.60 \pm 0.00$ &     $1.59 \pm 0.00$ &     $1.53 \pm 0.00$ & $0.67 \pm 0.00$ & $0.72 \pm 0.00$ & $0.86 \pm 0.00$ &                    \\
156 &      559 &     $3.09 \pm 0.06$ &     $4.71 \pm 0.21$ &    $10.23 \pm 0.91$ & $0.87 \pm 0.03$ & $1.68 \pm 0.05$ & $4.45 \pm 0.21$ &                    \\
157 &      564 &     $7.22 \pm 0.63$ &     $6.66 \pm 0.51$ &     $4.74 \pm 0.39$ & $1.80 \pm 0.10$ & $2.79 \pm 0.07$ & $6.17 \pm 0.20$ &                    \\
158 &      578 &     $1.63 \pm 0.01$ &     $1.66 \pm 0.01$ &     $1.77 \pm 0.04$ & $0.88 \pm 0.02$ & $0.96 \pm 0.02$ & $1.23 \pm 0.07$ &                    \\
160 &     2341 &     $2.39 \pm 0.10$ &     $2.22 \pm 0.07$ &     $1.62 \pm 0.01$ & $0.55 \pm 0.06$ & $0.61 \pm 0.04$ & $0.82 \pm 0.01$ &                    \\
161 &      576 &     $4.12 \pm 4.55$ &     $5.27 \pm 3.70$ &     $9.19 \pm 4.56$ & $3.58 \pm 2.60$ & $3.59 \pm 2.03$ & $3.61 \pm 1.10$ &                    \\
186 &     1135 &     $1.29 \pm 0.03$ &     $1.29 \pm 0.02$ &     $1.27 \pm 0.01$ & $2.25 \pm 0.06$ & $2.63 \pm 0.04$ & $3.92 \pm 0.04$ &                0.0 \\
187 &     1032 &     $2.15 \pm 0.04$ &     $2.26 \pm 0.03$ &     $2.64 \pm 0.01$ & $2.12 \pm 0.04$ & $2.33 \pm 0.03$ & $3.03 \pm 0.02$ &                0.0 \\
188 &     2103 &     $1.69 \pm 0.84$ &     $1.67 \pm 0.66$ &     $1.62 \pm 0.51$ & $7.69 \pm 1.99$ & $6.46 \pm 1.55$ & $2.28 \pm 0.75$ &                0.0 \\
189 &     2848 &     $2.09 \pm 0.01$ &     $2.01 \pm 0.01$ &     $1.74 \pm 0.01$ & $1.78 \pm 0.01$ & $1.51 \pm 0.01$ & $0.57 \pm 0.01$ &                    \\
190 &      909 &     $1.00 \pm 0.03$ &     $1.13 \pm 0.02$ &     $1.57 \pm 0.03$ & $2.86 \pm 0.08$ & $2.90 \pm 0.06$ & $3.06 \pm 0.06$ &                    \\
191 &     3317 &     $2.32 \pm 0.04$ &     $2.20 \pm 0.03$ &     $1.81 \pm 0.02$ & $1.53 \pm 0.02$ & $1.32 \pm 0.02$ & $0.62 \pm 0.02$ &                    \\
192 &      429 &     $1.71 \pm 0.06$ &     $1.62 \pm 0.05$ &     $1.33 \pm 0.02$ & $2.06 \pm 0.07$ & $1.90 \pm 0.06$ & $1.37 \pm 0.03$ &                    \\
193 &     3088 &     $2.95 \pm 0.11$ &     $2.79 \pm 0.09$ &     $2.24 \pm 0.04$ & $0.67 \pm 0.05$ & $0.74 \pm 0.04$ & $0.99 \pm 0.03$ &                    \\
195 &     2881 &     $2.42 \pm 0.05$ &     $2.39 \pm 0.05$ &     $2.29 \pm 0.04$ & $0.35 \pm 0.03$ & $0.40 \pm 0.03$ & $0.59 \pm 0.03$ &                    \\
196 &      474 &     $1.94 \pm 0.05$ &     $2.05 \pm 0.04$ &     $2.42 \pm 0.02$ & $2.93 \pm 0.06$ & $3.30 \pm 0.05$ & $4.56 \pm 0.03$ &                    \\
198 &     1003 &     $1.86 \pm 0.08$ &     $1.79 \pm 0.06$ &     $1.53 \pm 0.02$ & $1.57 \pm 0.07$ & $1.72 \pm 0.06$ & $2.22 \pm 0.03$ &                    \\
200 &     2037 &     $1.87 \pm 0.06$ &     $1.77 \pm 0.05$ &     $1.44 \pm 0.03$ & $0.78 \pm 0.05$ & $0.78 \pm 0.04$ & $0.77 \pm 0.05$ &                    \\
201 &      629 &     $3.11 \pm 0.10$ &     $3.19 \pm 0.08$ &     $3.48 \pm 0.03$ & $1.79 \pm 0.06$ & $2.06 \pm 0.04$ & $2.98 \pm 0.03$ &                    \\
202 &     3353 &     $4.02 \pm 0.40$ &     $3.60 \pm 0.31$ &     $2.17 \pm 0.05$ & $3.11 \pm 0.20$ & $2.69 \pm 0.16$ & $1.26 \pm 0.06$ &                    \\
203 &     1129 &     $0.85 \pm 0.02$ &     $0.87 \pm 0.02$ &     $0.94 \pm 0.01$ & $1.26 \pm 0.06$ & $1.21 \pm 0.04$ & $1.05 \pm 0.03$ &                    \\
204 &      631 &     $1.80 \pm 0.01$ &     $1.83 \pm 0.01$ &     $1.92 \pm 0.01$ & $0.27 \pm 0.01$ & $0.30 \pm 0.01$ & $0.41 \pm 0.01$ &                    \\
215 &     1451 &     $1.10 \pm 0.05$ &     $1.25 \pm 0.05$ &     $1.78 \pm 0.11$ & $0.51 \pm 0.07$ & $0.79 \pm 0.07$ & $1.74 \pm 0.17$ &                    \\
218 &      262 &     $0.16 \pm 0.00$ &     $0.18 \pm 0.01$ &     $0.26 \pm 0.05$ & $1.18 \pm 0.10$ & $1.96 \pm 0.57$ & $4.59 \pm 2.53$ &                    \\
219 &     3097 &     $0.52 \pm 0.04$ &     $0.63 \pm 0.05$ &     $1.02 \pm 0.18$ & $0.93 \pm 0.32$ & $1.03 \pm 0.31$ & $1.37 \pm 0.98$ &                    \\
227 &     2697 &     $0.54 \pm 0.04$ &     $0.62 \pm 0.03$ &     $0.86 \pm 0.04$ & $0.42 \pm 0.14$ & $0.44 \pm 0.11$ & $0.49 \pm 0.11$ &                    \\
\end{longtable}

\vspace{1ex}
\noindent\footnotesize\textbf{Note.} ID is the unique ID of the Spiderweb protocluster member. Effective radii and Sérsic indices are given with their symmetrical errors as $x \pm \Delta x$. Effective radii are in kpc. $\Sigma_{\rm SFR}$ is estimated using $R_{\rm eff,F182M}$ and is gien in $M_\odot\,\mathrm{yr}^{-1}\,\mathrm{kpc}^{-2}$ (only galaxies with measured SFR).
\end{center}

\end{appendix}

\end{document}